\documentclass[12pt, a4paper, notitlepage, superscriptaddress, floatfix, aps, pra]{revtex4-1}
\pdfoutput=1
\usepackage[usenames]{color}
\usepackage{graphicx}
\usepackage[pdftex,unicode,colorlinks,citecolor=blue,linkcolor=blue]{hyperref}

\begin{document}
\title{Quantum expander for gravitational-wave observatories}

\author{Mikhail Korobko}
\affiliation{Institut f\"ur Laserphysik und Zentrum f\"ur Optische Quantentechnologien, Universit\"at Hamburg, Luruper Chaussee 149, 22761 Hamburg, Germany}
\email{mkorobko@physnet.uni-hamburg.de}
\author{Yiqiu Ma}
\affiliation{Theoretical Astrophysics 350-17, California Institute of Technology, Pasadena, California 91125, USA}
\author{Yanbei Chen}
\affiliation{Theoretical Astrophysics 350-17, California Institute of Technology, Pasadena, California 91125, USA}
\author{Roman Schnabel}
\affiliation{Institut f\"ur Laserphysik und Zentrum f\"ur Optische Quantentechnologien, Universit\"at Hamburg, Luruper Chaussee 149, 22761 Hamburg, Germany}

% Include the date command, but leave its argument blank.
%\date{}

% Make the title.

\begin{abstract}
Quantum uncertainty of laser light limits the sensitivity of gravitational- wave observatories. In the past 30 years, techniques for squeezing the quantum uncertainty as well as for enhancing the gravitational-wave signal with optical resonators were invented. Resonators, however, have finite linewidths; and the high signal frequencies that are produced during the scientifically highly interesting ring-down of astrophysical compact-binary mergers cannot be resolved today. Here, we propose an optical approach for expanding the detection bandwidth. It uses quantum uncertainty squeezing inside one of the optical resonators, compensating for finite resonators’ linewidths while maintaining the low-frequency sensitivity unchanged. Introducing the quantum expander for boosting the sensitivity of future gravitational-wave detectors, we envision it to become a new tool in other cavity-enhanced metrological experiments.\\
\end{abstract}

\maketitle

% In setting up this template for *Science* papers, we've used both
% the \section* command and the \paragraph* command for topical
% divisions.  Which you use will of course depend on the type of paper
% you're writing.  Review Articles tend to have displayed headings, for
% which \section* is more appropriate; Research Articles, when they have
% formal topical divisions at all, tend to signal them with bold text
% that runs into the paragraph, for which \paragraph* is the right
% choice.  Either way, use the asterisk (*) modifier, as shown, to
% suppress numbering.

\section{Introduction}
The dawn of gravitational-wave astronomy has begun with the historic detection of binary black hole coalescence in 2015~\cite{Abbott2016d}, and several more detections that followed in the years after~\cite{ligo2018gwtc}.
The latest observation of gravitational waves was from a binary neutron star inspiral.
It was succeeded by observations of a broad spectrum of electromagnetic counterparts~\cite{TheLIGOScientificCollaboration2017a,TheLIGOScientificCollaboration2017b} and demonstrated that gravitational-wave astronomy is invaluable for understanding the Universe\,~\cite{TheLIGOScientificCollaboration2017c,TheLIGOScientificCollaboration2017d}.
Further increasing the sensitivity of GWOs is of utmost importance to maximize the scientific output of combined multi-messenger astronomical observations.

\begin{figure}[t]
	\includegraphics{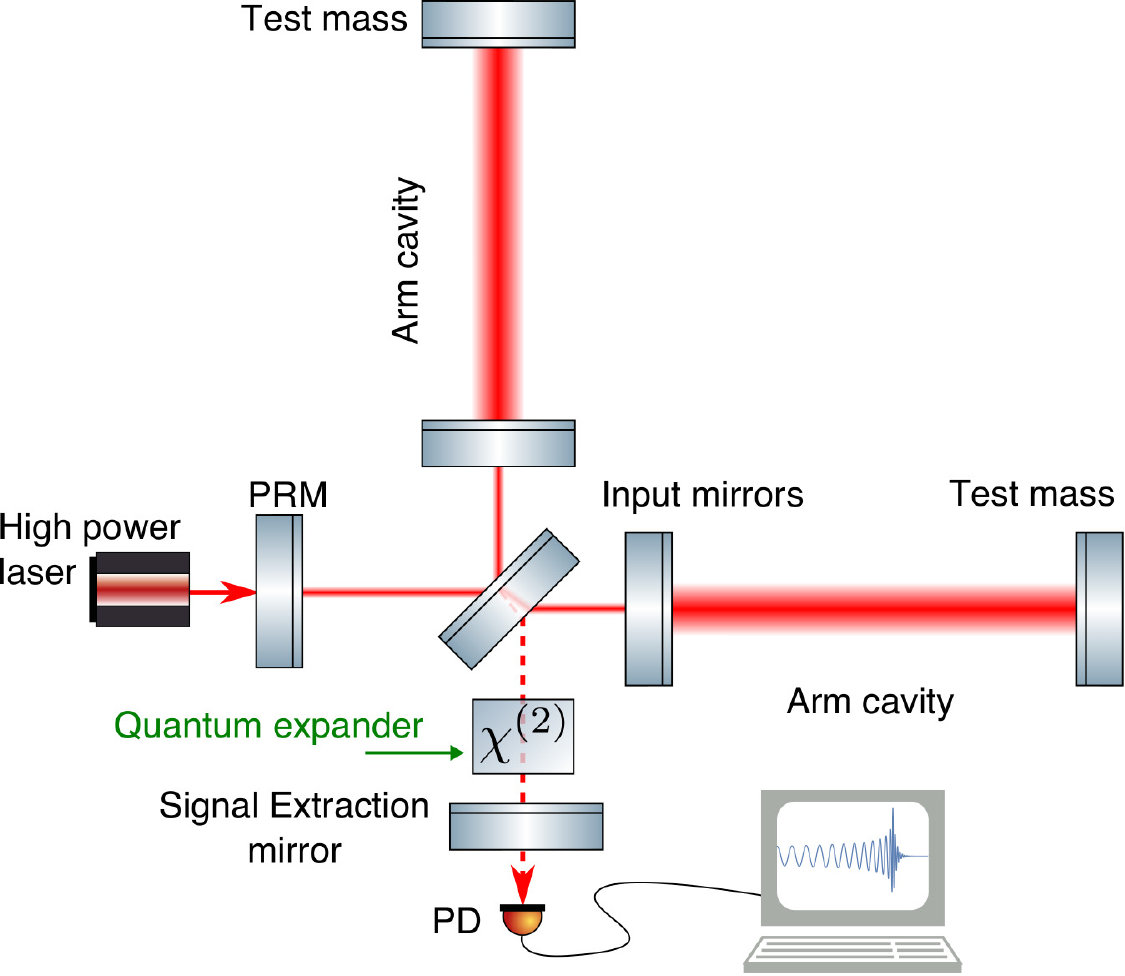}
	\caption{Conceptual representation of the GW observatory with our quantum expander. The relative change in the distance between the central beam splitter and the test masses due to a gravitational wave is measured on the signal port with a photodiode PD. Optical cavities in the arms are used to enhance the light power and the signal. Additional mirrors independently enhance the signal (signal extraction mirror) and power (power recycling mirror, PRM). The external squeezed light field is injected into a dark port to suppress the shot noise. We add a nonlinear $\chi^{(2)}$ crystal in the signal extraction cavity, formed by the SE mirror and input mirrors, which creates internally squeezed light field to boost the high-frequency sensitivity.}\label{fig:setup}
\end{figure}

Gravitational-wave observatories, such as Advanced LIGO~\cite{Aasi2015}, Advanced Virgo~\cite{Acernese2015}, KAGRA~\cite{Aso2013} and GEO600~\cite{Smith2004}, are based on the Michelson interferometer topology (see Fig.\,\ref{fig:setup}), where the incoming gravitational wave changes the relative optical path length of two interferometer arms.
The ability of the observatories to measure gravitational waves is limited by various disturbances that also change the differential path length or manifest themselves as such.
The main noise source at signal frequencies above $\sim 50$\,Hz  in the current generation of GW observatories is the quantum uncertainty of the light field, which results in shot noise (photon counting noise)~\cite{Aasi2015,Martynov2016}.
Noise at lower frequencies has contributions of several origins such as Brownian motion of the mirror surfaces and suspensions, as well as the quantum radiation pressure noise, which comes from mirrors' random motion due to quantum fluctuations of light power~\cite{Caves1980a,Braginsky1964}.
All these noise sources give contribution to the photocurrent of the photodiode placed on the signal port of the detector.
The observatory's sensitivity to the GW signal, \emph{i.e.} its ability to discriminate between the GW signal and noise, is given by the observatory's signal-to-noise ratio (SNR).
The sensitivity ultimately is limited by the ``quantum Cramer-Rao bound'' (QCRB)~\cite{Tsang2011} of the detector.
For continuous signals the best sensitivity at each frequency is determined by the radiation pressure force exerted on the test mass by quantum fluctuations of the light field~\cite{Miao2016}.
One way to lower the QCRB is to increase the quantum uncertainty in the amplitude of the light field by injecting phase-squeezed vacuum states of light into the interferometer~\cite{Yuen1976,Caves1980a,Schnabel2017}, which has become a well-established technique for GW observatories~\cite{Abadie2011,Abadie2013,Grote2013}

A conventional way to increase the signal response of the detector is to use the optical resonators, as implemented already in the first generation of GWOs~\cite{Drever1983a}.
The resonance buildup of optical energy in resonators increases the radiation pressure force~\cite{00p1BrGoKhTh} and hence lowers the QCRB.
The current (second) generation design includes Fabry-Perot cavities in the arms, as well as the \emph{signal-extraction} (SE) cavity on the dark port of the detector~\cite{Meers1988}.
Resonators, however, only significantly lower the QCRB at frequencies below the resonator’s linewidth, \emph{i.e.} they reduce the observatory's detection bandwidth~\cite{Mizuno1995a}.
Injection of squeezed states, mentioned above, is not able to counteract the loss of bandwidth due to resonators.

The issue of detector \emph{bandwidth} becomes crucial in the era of multi-messenger astronomy~\cite{TheLIGOScientificCollaboration2017a}.
The information about physics of extremal nuclear matter is hidden in waveforms of gravitational waves radiated from the post-merger remnants of binary neutron star systems\,~\cite{Faber2012}.
Obtaining this information is important for unraveling the physics of compact astrophysical objects --- the engines that drive gamma-ray bursts, the origin of heavy elements and possible modifications to general relativity~\cite{Baiotti2017,Conklin2017}.
These waveforms have typical frequencies above 1\,kHz, where the sensitivity of current observatories degrades due to the bandwidth.

Over the past 20 years the challenge of increasing the bandwidth without changing the peak sensitivity at low frequencies has become one of the cornerstones for the design of future gravitational-wave detectors~\cite{Wicht1997,Pati2007}.
Previous concepts involved use of unstable optomechanical or atomic systems in so-called ``negative dispersion'' operation~\cite{Yum2013,Zhou2015,Miao2105,Qin2015,Miao2015}.
In this work we propose a new and all-optical concept without instabilities, which targets on achieving the same goal, \textit{i.e.} arbitrary expansion of detection bandwidth, given low enough quantum decoherence.
This \textit{quantum-expanded signal extraction} concept is based on optical parametric amplification process inside the interferometer, which allows to increase the quantum fluctuations in the amplitude of the light by introducing quantum correlations, thereby reducing the QCRB.
Due to the optical coupling between the cavities, the quantum uncertainty at high frequencies gets squeezed such that it compensates the reduction in signal enhancement due to the cavity linewidth.
At low frequencies neither signal nor quantum noise change, which maintains the existing sensitivity, which is optimized for observing the pre-merger stages of binary coalescence.
Our approach is fully compatible with further enhancements to the detector design, such as injection of frequency-dependent squeezed light or variational readout~\cite{Kimble2000,Chelkowski2007a,Danilishin2012,Miao2014}.

Placing an optical parametric amplifier inside the detector has been considered for other purposes before~\cite{Rehbein2005,Somiya2016,Korobko2017a,Korobko2017b}, and all-optical quantum expansion of bandwidth has never been proposed so far.

\section{Quantum boost of high-frequency sensitivity}
In future, GW interferometers will operate with the signal port being at the dark fringe.
In this operating condition all of the light power pumped into the interferometer is reflected towards the source of the pump light.
The only light that leaves the interferometer through the dark signal port corresponds to the signal caused by the dynamical change in the differential arm length, \emph{e.g.} due to a GW.
The zero-point fluctuation that enters the dark port defines the shot noise of the interferometer.

\begin{figure}[h!]
        \includegraphics{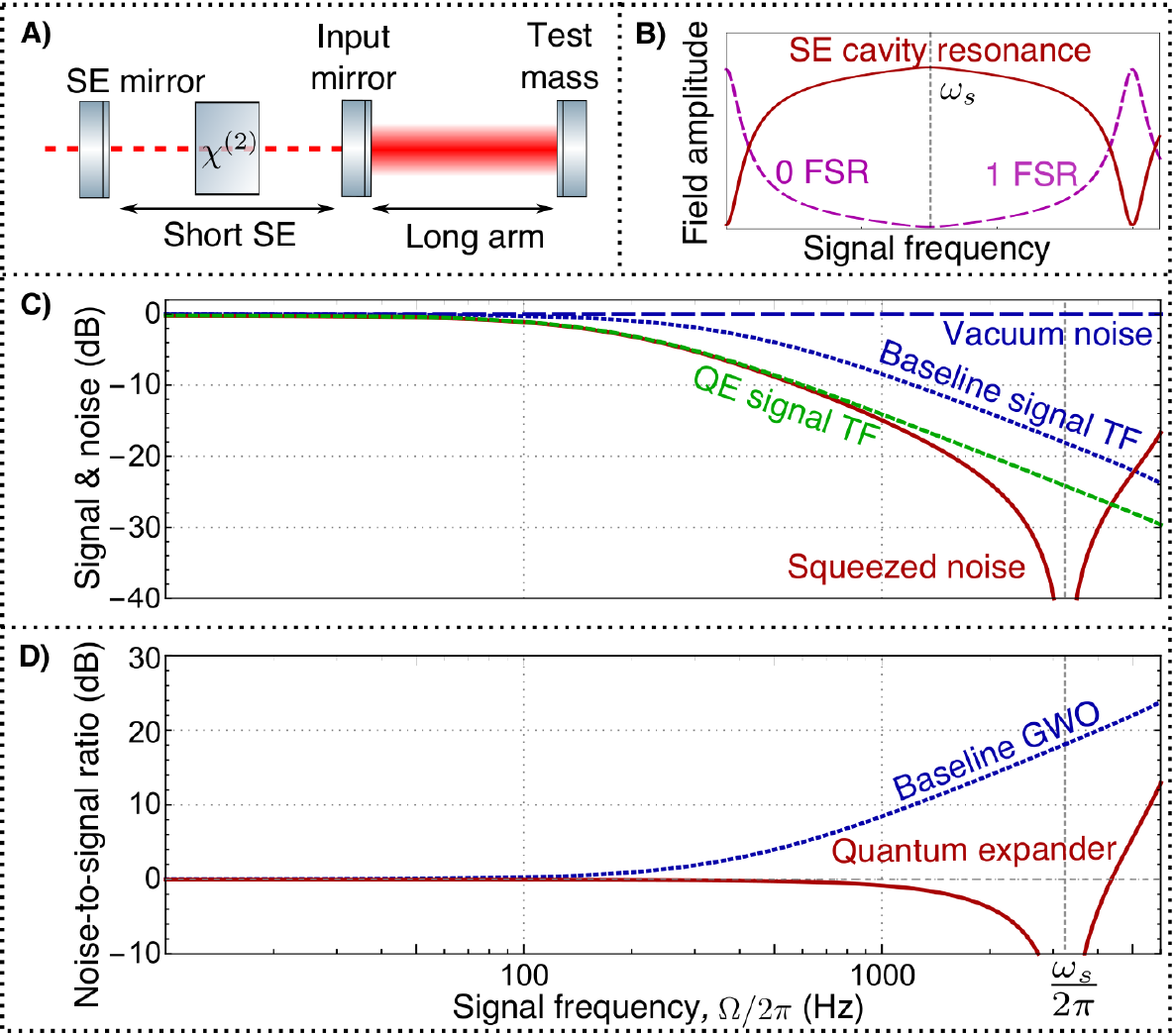}
        \caption{Concept of the quantum expander.
          A) Model system of two coupled cavities, arm and signal extraction (SE), with nonlinear crystal inside SE cavity; B) resonance enhancement of the SE mode at frequencies close to $\omega_s$ and suppression at low frequencies, with two free spectral ranges (FSR) of the arm cavity; C) suppression of the shot noise at high frequency by the quantum expander (red) compared to the vacuum level (blue), in comparison to the scaling of the signal transfer function (TF) due to the cavity linewidth with quantum expander (green) and without (blue), where the signal is suppressed by 6\,dB due to the parametric process; D) noise-to-signal ratio for the detector with quantum expander (red) and without (blue). On C) the quantum expander noise squeezing has exactly the same scaling as signal reduction due to the cavity bandwidth, so the signal-to-noise ratio is boosted at high frequencies, as seen on D).}
\label{fig:signal_noise}
\end{figure}

With respect to the quantum noise and the signal, the interferometer topology can be conceptually represented by a simpler system of two coupled cavities~\cite{Buonanno2003}: the arm cavity with optical mode $\hat{a}$, and the signal-extraction cavity, formed by the front mirror of the arm cavity and the signal-extraction mirror, with optical mode $\hat{a}_q$, see Fig.\,\ref{fig:signal_noise}A.
The two modes are coupled through the partially reflective front mirror of the arm cavity, with a coupling frequency $\omega_s$, which depends on the reflectivity of this front mirror.
For illustrative purposes we limit the discussion to the interaction of these two modes, while the complete description should include the effects of the next free spectral range of the arm cavity and the interaction with its mode.
In this assumption the system can be described by a standard Hamiltonian for coupled harmonic oscillators: $\hat{H}/\hbar =  \omega_0 \hat{a}^\dag \hat{a} + \omega_0 \hat{a}_q^\dag \hat{a}_q + \omega_s (\hat{a}_q^\dag \hat{a}+\hat{a}^\dag \hat{a}_q)$.
If the system is excited at one of the normal frequencies, $\omega_0 \pm \omega_s$, the excitation energy is equally split between the two modes.
However, when one of the modes (\emph{e.g.} $\hat{a}_q$) is excited at $\omega_0$, the complete energy gets redistributed into the coupled mode ($\hat{a}$).
In this way when mode $\hat{a}_q$ is open to the environment and driven by the incoming zero-point fluctuation, its noise components are strongly suppressed at sideband frequencies $\omega_0 \pm \Omega, \Omega\ll\omega_s$, and all the energy at these frequencies goes into the coupled mode $\hat{a}$.
For larger frequency $\Omega$, the noise becomes resonant inside the SE cavity as well, and reaches its maximum at $\omega_s$, as can be seen of Fig.\ref{fig:signal_noise}B.
It is this particular resonant structure that we take advantage of for boosting the sensitivity of the detector at high frequencies.
We propose to place an optical parametric amplifier,  \textit{e.g.} a $\chi^{(2)}$ nonlinear crystal, inside the SE cavity.
The parametric process will amplify the fluctuations in one quadrature of the mode $\hat{a}_q$, and suppress the fluctuations in its conjugate counterpart.
Depending on the sideband frequency $\Omega$, the amplification strength varies due to the presence of the coupled cavity structure.
At frequencies around $\omega_0$, the excitation of mode $\hat{a}_q$ is suppressed, so the parametric process is inefficient, and no squeezing is produced.
At the same time, the SE cavity is resonant for higher frequencies $\Omega \sim \omega_s$, so the crystal produces a high squeeze factor.
The suppression of shot noise at the frequencies $0\ll\Omega\ll\omega_s$ happens exactly at the same rate as the reduction in the signal amplification due to the detector bandwidth, see Fig.\,\ref{fig:signal_noise}C.
The two processes compensate each other, and the signal-to-noise ratio remains constant, thus the bandwidth is expanded, see Fig.\,\ref{fig:signal_noise}D.

Quantum expansion effect can be demonstrated in more detail by formulating a complete Hamiltonian of the model two-mode system (for a general analysis of the system, see the Supplementary Material):
\begin{eqnarray}
&\hat{H}& = \hat{H}_0 + \hat{H}_{\rm int} + \hat{H}_\gamma + \hat{H}_{x} - F_{\rm GW} x; \\
&\hat{H}_0& = \hbar \omega_0 \hat{a}^\dag \hat{a} + \hbar \omega_0 \hat{a}_q^\dag \hat{a}_q;\\
&\hat{H}_{\rm int}& =  \hbar \omega_s \hat{a}_q^\dag \hat{a} + \frac{1}{2}\hbar \kappa \beta e^{-2i\omega_0 t} \hat{a}_q^\dag \hat{a}_q^\dag e^{i\phi} + h.c.;\\
&\hat{H}_\gamma& = i \hbar \sqrt{2\gamma} \int_{-\infty}^{\infty} \left(\hat{a}_q^\dag(\omega)\hat{a}_{\rm in}(\omega) - \hat{a}_{\rm in}^\dag(\omega)\hat{a}_q(\omega)\right)d\omega;\\
&\hat{H}_x& = - \hat{F}_{\rm rp} \hat{x} = -\hbar G_0 \hat{a}^\dag \hat{a} \hat{x},
\end{eqnarray}
where $\hat{a}, \hat{a}_q$ are the arm cavity and SE cavity modes, and $\omega_0$ is their natural resonance frequency; $\omega_s = c\sqrt{T_{\rm ITM}/(4 L_{\rm SE} L_{\rm arm})}$ is the coupling rate between two cavities, $T_{\rm ITM}$ is the transmission of the front mirror of the arm cavity, $L_{\rm SE}, L_{\rm arm}$ are the lengths of the signal extraction and arm cavity, respectively;  $\gamma = c T_{\rm SE}/(4 L_{\rm SE})$ is the coupling rate of the SE mode to the continuum of input modes $\hat{a}_{\rm in}$; $x$ is the displacement of the test mass partially in reaction to the gravitational-wave tidal force $F_{\rm GW}$; the mirror motion $x$ is coupled via the radiation-pressure force $\hat{F}_{\rm rp}$ to the cavity mode with strength $G_0=\omega_0/L_{\rm arm}$; $\kappa$ is the coupling strength due to a crystal nonlinearity under a second harmonic pump field with amplitude $\beta$ and phase $\phi$.
The pump field is assumed to be classical and its depletion is neglected.
Quantum expansion affects only the high frequency sensitivity, which is dominated by the shot noise.
This justifies us to ignore in this simple model the effects of the quantum radiation pressure on the dynamics of the test mass, effectively assuming infinite mass of the mirrors, whose displacement is caused only by the GW strain $h_0 = x/L_{\rm arm}$.
Note that the expression for the coupling frequency $\omega_s$ is modified when the higher FSR of the arm cavity is taken into account, and in the current form only applicable when $\omega_s\ll \omega_{\rm FSR} = c/(2L_{\rm arm})$.

The light field in the system can be expressed in terms of the input fields by solving the Hamiltonian above.
We write the input-output relations for the amplitude and phase quadrature of the light (denoted by upper indices $(1,2)$ correspondingly), representing the field leaving the detector $\hat{a}_{\rm out}^{(1)}$, field inside the SE cavity $\hat{a}_{q}^{(2)}$, and field inside the arm cavity $\hat{a}^{(1)}$ in terms of input noise fields $\hat{a}_{\rm in}^{(1,2)}$:
\begin{eqnarray}
  &\hat{a}_{\rm out}^{(1)}(\Omega) &= \hat{a}_{\rm in}^{(1)}(\Omega)\frac{(\gamma - \chi)\Omega + i(\Omega^2-\omega_s^2)}{(\gamma + \chi )\Omega - i(\Omega^2-\omega_s^2)} +  h_0(\Omega)\frac{2iG\sqrt{\gamma}\omega_s}{(\gamma + \chi)\Omega - i(\Omega^2-\omega_s^2)} ,\\
  &\hat{a}_q^{(2)}(\Omega) &= \hat{a}_{\rm in}^{(2)}(\Omega)\frac{\sqrt{2\gamma}\Omega}{(\gamma + \chi)\Omega - i(\Omega^2-\omega_s^2)} + h_0(\Omega)\frac{iG\omega_s}{(\gamma + \chi)\Omega - i(\Omega^2-\omega_s^2)},\\
  &\hat{a}^{(1)}(\Omega) &= \hat{a}_{\rm in}^{(2)}(\Omega)\frac{i\sqrt{2\gamma}\omega_s}{(\gamma - \chi )\Omega - i(\Omega^2-\omega_s^2)},
\end{eqnarray}
where we linearized the system dynamics and introduced an effective parametric gain $\chi = \kappa \beta$, effective signal coupling strength $G = \sqrt{2 P_c L_{\rm arm}\omega_0/(\hbar c)}$ and optical power inside the arm cavity $P_c = \hbar \omega_0 \bar{a}$, with $\bar{a}$ being an average amplitude of the mode $\hat{a}$.
Several features can be seen in these equations.
First, when we remove the crystal (\emph{i.e.} $\chi=0$) in the typical operational range of GW observatories $\Omega\ll\omega_s$, the input-output relation Eq.(6) reduces to the standard one for a baseline GWO~\cite{Buonanno2003,martynov2019}, with the detection bandwidth given by: $\gamma_{\rm baseline}= \omega_s^2/\gamma = c T_{\rm ITM}/(T_{\rm SE}L_{\rm arm})$.
Second, the noise term in Eq.(7) is strongly suppressed at zero sideband frequency, as we described above in the example with two coupled modes: $\hat{a}_q^{(2)}(0) = h(0) G/\omega_s$, therefore virtually no squeezing is produced at low frequencies.
The noise on the output in Eq.(6) at low frequencies is defined by the vacuum field reflected directly off the signal extraction mirror.
Third, when the sideband frequency matches the normal mode frequency, $\Omega = \omega_s$, the signal mode takes the form: $\hat{a}_{\rm out}^{(1)}(\omega_s) = \hat{a}_{\rm in}^{(1)}(\gamma - \chi)/(\gamma + \chi) + 2ih_0(\omega_s)G\sqrt{\gamma}/(\gamma + \chi)$.
This equation shows that when the parametric gain approaches threshold: $\chi\rightarrow \gamma$, the noise term becomes almost infinitely squeezed~\cite{Schnabel2017}, but signal gets deamplified by maximally a factor of 2.
Despite the signal deamplification, ideally the SNR in this case can become infinite, as we show below by computing the sensitivity of the quantum-expanded detector.

The noise spectral density of the GWO with quantum expander, normalized to the unity of strain $h$, can be obtained from Eqs.(6)-(8) and further approximated in the typical regime where GWOs operate, $\gamma\gg\omega_s\gg\Omega$, as following:
\begin{equation}
	S_{h}(\Omega) = \frac{\hbar c}{8 \omega_0 L_{\rm arm} P_c}\frac{(\Omega^2-\omega_s^2)^2 + (\gamma-\chi)^2\Omega^2}{\gamma \omega_s^2} \approx \frac{\hbar c}{8 \omega_0 L_{\rm arm} P_c}\frac{\gamma_q^2 + \Omega^2}{\gamma \omega_s^2}(\gamma-\chi)^2,
\end{equation}
with the new detection bandwidth defined as $\gamma_{q} = \omega_s^2/(\gamma-\chi)$.
Without the quantum expansion, $\chi=0$, the baseline sensitivity decreases with the frequency increase, limited by the detector's bandwidth $\gamma_{\rm baseline} = \omega_s^2/\gamma$:
\begin{equation}
	S_{h}^{\rm baseline}(\Omega) = \frac{\hbar c}{8 \omega_0 L_{\rm arm} P_c}\frac{(\Omega^2-\omega_s^2)^2 + \gamma^2\Omega^2}{\gamma \omega_s^2} \approx \frac{\hbar c}{8 \omega_0 L_{\rm arm} P_c}\frac{\gamma_{\rm baseline}^2 + \Omega^2}{\gamma \omega_s^2}\gamma^2.
\end{equation}
The detection bandwidth $\gamma_q$ can ideally be expanded infinitely (in the two-mode approximation) by a factor of $\gamma/(\gamma-\chi)\rightarrow\infty$ when squeezing approaches the threshold point $\chi=\gamma$.
At this point the sensitivity is given by
\begin{equation}
	S_{h}(\Omega) = \frac{\hbar c}{8 \omega_0 L_{\rm arm} P_c}\frac{\omega_s^2}{\gamma},
\end{equation}
which is approximately frequency independent under $\Omega\ll\omega_s$, as a result of expanded bandwidth $\gamma_q$.
In reality, even in the lossless case, the bandwidth is still limited by the next free spectral range of the arm cavity, and the detector's response function to the gravitational wave (which becomes important when the gravitational wavelength is comparable to the arm length).
\begin{figure}[t!]
        \includegraphics{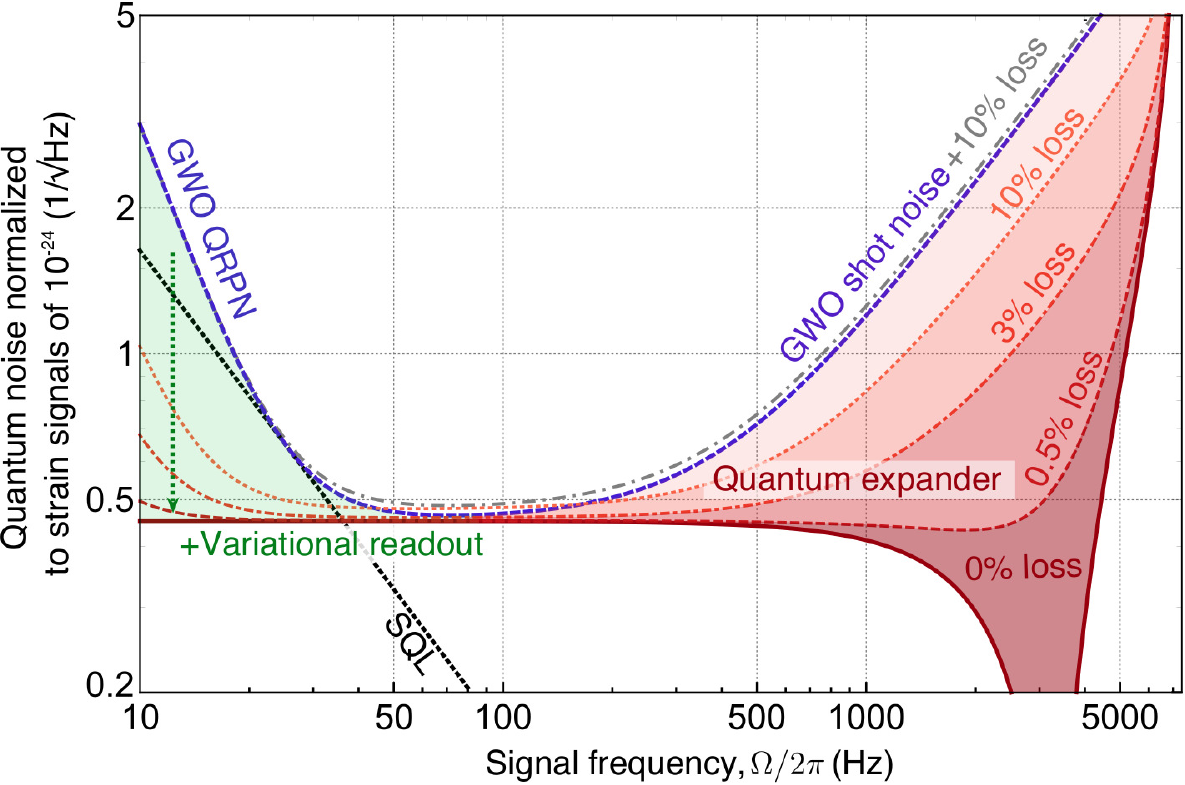}
        \caption{Effect of the quantum expander on the detector's sensitivity to gravitational-wave strain $S_h(f)$ (red), in combination with variational readout~\cite{Kimble2000}.
				The photon shot-noise limited bandwidth of the semiclassical Gravitational Wave Observatory (GWO, blue dashed line) is expanded by squeezing operation inside the detector at high frequencies (solid red line, red shading).
				The effect reduces as quantum decoherence due to optical loss is introduced (different shades of red for quantum expander, gray dot-dashed line for semiclassical GWO).
				At low frequencies quantum noise remains unaffected by quantum expansion, and allows to use the variational readout (green shading) to evade the quantum radiation-pressure noise (QRPN).
       The boundary where the QRPN becomes equal to the shot noise at different light powers, know as the Standard Quantum Limit (SQL) is plotted in black dots.
			 The parameters used for plotting are based on the benchmark parameter set for the 3d generation of GWOs: optical wavelength $\lambda = 1550$\,nm; light power inside the arm cavity $P_c = 4$\,MW; arm cavity length $L=20$\,km; SE cavity length $l_s=56$\,m; mirror mass $m=200$\,kg; input mirror power transmission $T_i=0.07$; SE mirror power transmission $T_s=0.35$.}
\label{fig:wlc_noloss}
\end{figure}
The effect of the quantum expander on a baseline GW observatory is shown on Fig.\,\ref{fig:wlc_noloss}.
To produce this figure we compute the sensitivity based on the transfer matrix approach (as presented in the Supplementary Material), which better describes the high-frequency behavior in the longer detectors, i.e. when $\omega_s\sim\omega_{FSR}$.
It also takes into account the effects of quantum radiation pressure noise, quantum decoherence, the next free spectral ranges of the cavities as well as the response function of the detector to gravitational waves.

The sensitivity of any gravitational-wave observatory is ultimately limited by its quantum Cramer-Rao bound (QCRB) $S_{h}^{\rm QCRB}(\Omega)$~\cite{Miao2016}.
The conditions for reaching its quantum Cramer-Rao bound are that (i) the quantum radiation pressure noise is evaded, and (ii) the upper and lower optical sidebands generated by the GW are equal in amplitude.
The quantum expander configuration does not affect the QRPN, and allows to satisfy condition (i) at low frequency by back-action evading techniques (\emph{e.g.} variational readout).
We prove that the condition (ii) is satisfied by directly computing the QCRB in the case of GW detectors is defined as follows~\cite{Miao2016}:
\begin{equation}
  S_{h}^{\rm QCRB}(\Omega) = \frac{\hbar^2}{2 L_{\rm arm}^2S_{FF}(\Omega)} = \frac{\hbar c}{4 \omega_0 L_{\rm arm} P_c}\frac{1}{S_{aa}(\Omega)},
\end{equation}
where $S_{FF}(\Omega)$ is the single-sided spectrum of the radiation-pressure force $\hat{F}_{\rm rp}$, and $S_{aa}(\Omega)$ is the noise spectrum of the arm cavity field, which one can compute from Eq.~9:
\begin{equation}
  S_{aa}(\Omega) = \frac{2 \gamma \omega_s^2}{(\gamma-\chi)^2\Omega^2 + (\Omega^2-\omega_s^2)^2}
\end{equation}
Therefore the limit on the sensitivity is given by the QCRB in the following form:
\begin{equation}
  S_{x}^{\rm QCRB}(\Omega)= \frac{\hbar c}{4 \omega_0 L_{\rm arm} P_c}\frac{(\Omega^2-\omega_s^2)^2 + (\gamma-\chi)^2\Omega^2}{2 \gamma \omega_s^2},
\end{equation}
which is identical to Eq.\,(10).
The sensitivity becomes unbounded (QCRB turns to zero) at the parametric threshold $\chi = \gamma$ at frequency $\Omega = \omega_s$.

This calculation demonstrates, that in the ideal case our quantum expander operates exactly at the QCRB, which on top is strongly reduced at high frequencies compared to the baseline GWO.

\section{Discussion and outlook}

All observatories of the current generation are going to operate with external-squeezing injection soon.
Quantum-expanded signal extraction will further reduce the shot noise at high frequencies, without affecting the established improvement factor from external squeezing.
The quantum noise at \emph{low} frequencies (QRPN) will remain unchanged.
This differs our approach from other designs targeting the high-frequency sensitivity~\cite{Corbitt2004,Miao2017,martynov2019}.
The QRPN can be suppressed independently using already developed approaches using frequency dependent squeezing, variational readout or quantum non-demolition measurements~\cite{Kimble2000,Chelkowski2007a,Danilishin2012,Miao2014}.
In Fig.\,\ref{fig:wlc_noloss} we show combination of the quantum expander with variational readout.

\emph{Quantum decoherence}. Non-classical light is sensitive to decoherence, \emph{i.e.}\ to optical loss, which destroys the inherent quantum correlations~\cite{Grynberg2010}.
Losses occur inside the detector as well as on the readout, and have multiple contributions.
Any squeezed light application as well as QRPN suppression technique is limited by optical loss, and the proposed scheme is not an exception.
The quantum expander relies on squeezing operation inside the interferometer to compensate the loss of signal amplification due to the finite cavity linewidth.
The higher the squeeze factor is, the more it is susceptible to optical loss.
The effect of different readout loss is shown on Fig.\,\ref{fig:wlc_noloss}.
In the current generation of GWOs, the optical readout loss is on the order of 10\%~\cite{Oelker2014}, and in next observatory generation 3--5\% might be achievable~\cite{Schreiber2017}.
With advanced techniques, which have been proposed~\cite{Caves1981,Knyazev2018}, but yet to be explored experimentally, the readout loss could conceivably be reduced to be as small as 0.5\%.
Introducing a nonlinear crystal inside the detector will increase the internal loss, due to additional optical surfaces and optical absorption.
A more detailed discussion of different loss sources can be found in the Supplementary Material.
While we believe the added loss due to a crystal can be relatively small (see \emph{e.g.} the discussion in~\cite{Korobko2017a}), we consider our work to be a strong a motivation for detailed experimental research and development.

\begin{figure}[t!]
        \includegraphics{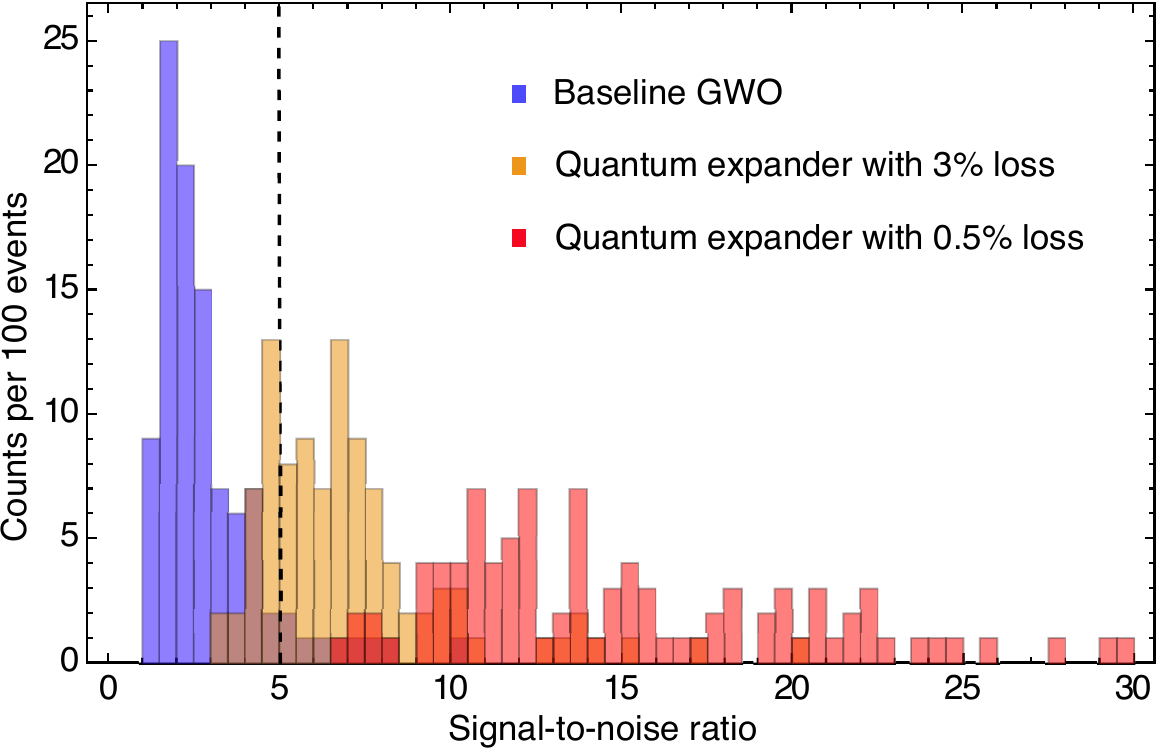}
        \caption{Histogram for signal-to-noise ratio of the loudest event for 100 realizations in the Monte-Carlo simulation. Blue bins represent the SNR of our baseline gravitational wave observatory. Orange and red bins are associated with the quantum expander with total loss around $3\%$ and $0.5\%$, respectively.
        The black dashed line indicates a detection threshold (${\rm SNR}=5$).
				We used the equation of state in~\cite{Janka2011,Yang2018} and the binary merger rate is taken to be $R=1.54{\rm Mpc}^{-3}{\rm Myr}^{−1}$.
				The mass distribution for each neutron star in the binary is taken to be gaussian centered around 1.33 $M_{\odot}$.}
\label{fig:histogram}
\end{figure}

When the quantum expander is combined with external-squeezing injection, the overall squeeze factor increases further.
This makes the requirements for low optical loss more stringent.
The benefit from quantum expansion in combination with external squeezing depends not only on the amount of loss, but also on the places where it occurs.
There exists an optimal parametric gain in quantum expander that maximizes the sensitivity by balancing the signal deamplification in parametric process and squeeze factor~\cite{Korobko2017a}.
Ultimately every specific design of the detector has to be optimized with respect to optical parameters to be able to maximally benefit from quantum expansion.

We envision the quantum expander to become beneficial in future generations of GW observatories when the technological progress allows to lower the optical losses, and detectors become longer and overall more sensitive: \emph{e.g.} in the extensions of the third generation of observatories (Einstein Telescope and Cosmic Explorer) and beyond.

\emph{Astrophysical implications}.

Quantum expansion of the bandwidth has a high potential for allowing the observation of astrophysical objects at different stages of their evolution, starting from the pre-merger and finishing with the decaying oscillations of the newly formed object.
Such observations can give us a better understanding of physics of neutron stars.
The histogram in Fig.\,\ref{fig:histogram} shows the improvement of detectability of gravitational-wave signal emitted by binary neutron star merger remnants, according to the sensitivities in Fig.\,\ref{fig:wlc_noloss}.
There is about $9\%$ chance to have a single loud event surpassing the detection threshold after a full-year data acquisition in a baseline GWO, given a specific equation of state for the remnants of neutron star merger~\cite{Janka2011,Yang2018}.
With quantum expader the chance rises to roughly $76\%$ and almost $100\%$ for the system with $3\%$ and $0.5\%$ optical loss, respectively, which shows a significant improvement relative to the baseline configuration.

We anticipate that also other metrological~\cite{Szczykulska2016,Li2018} as well as optomechanical~\cite{Aspelmeyer2014a,Khalili2014} experiments can benefit from our approach of using a coupled-cavity system with a parametric amplifier inside for bandwidth expansion.

%%%%%%%%%%%%%%%%%%%%%%%%%%%%%%%%%%%%%%
\subsection*{Acknowledgments}
We thank Farid Khalili and Sebastian Steinlechner for valuable comments.
\textbf{Funding:} M.Korobko is supported by the Deutsche Forschungsgemeinschaft (DFG) (SCHN 757/6-1); R.Schnabel is supported by the European Research Council (ERC) Project “MassQ” (Grant No. 339897) and the Deutsche Forschungsgemeinschaft (DFG) (SCHN 757/6-1); Y. Chen and Y. Ma are supported by the National Science Foundation through Grants PHY-1708212 and PHY-1708213,the Brinson Foundation, and the Simons Foundation (Award Number 568762).  \\
\textbf{Author contributions:} M.Korobko: conceptualization, formal analysis, methodology, software, visualization, writing; Y.Ma: formal analysis, methodology, validation, software, writing (review \& editing);
Y.Chen: supervision, validation, writing (review \& editing);
R.Schnabel: project administration, supervision, writing (review \& editing).\\
\textbf{Competing interests:} authors declare no competing interests.\\
\textbf{Data and materials availability:} data and code used to produce the figures are available by request to the corresponding author.\\
\newpage
\appendix
% \section{Supplementary Text}
% In the supplementary text we discuss the experimental feasibility of the quantum expander, with emphasis on the optical loss sources and the possible benefit from the quantum expander in the presence of external squeezing; derive the equations in the main text based on the simplified input-output relations and on the Hamiltonian description of the system; give details on the astrophysical implications of use of quantum expander; provide details of derivation of the full spectral density of the detector with various imperfections based on the optical transfer matrix approach.
\section{Experimental feasibility}
In this section we discuss some of the issues of the experimental feasibility.
We indicate the main sources of loss and their contribution into the resulting sensitivity, and analyze the achievable benefit from quantum expansion when combined with external squeezed-light injection.

\subsection{Optical loss}
\begin{figure}[h!]
        \includegraphics{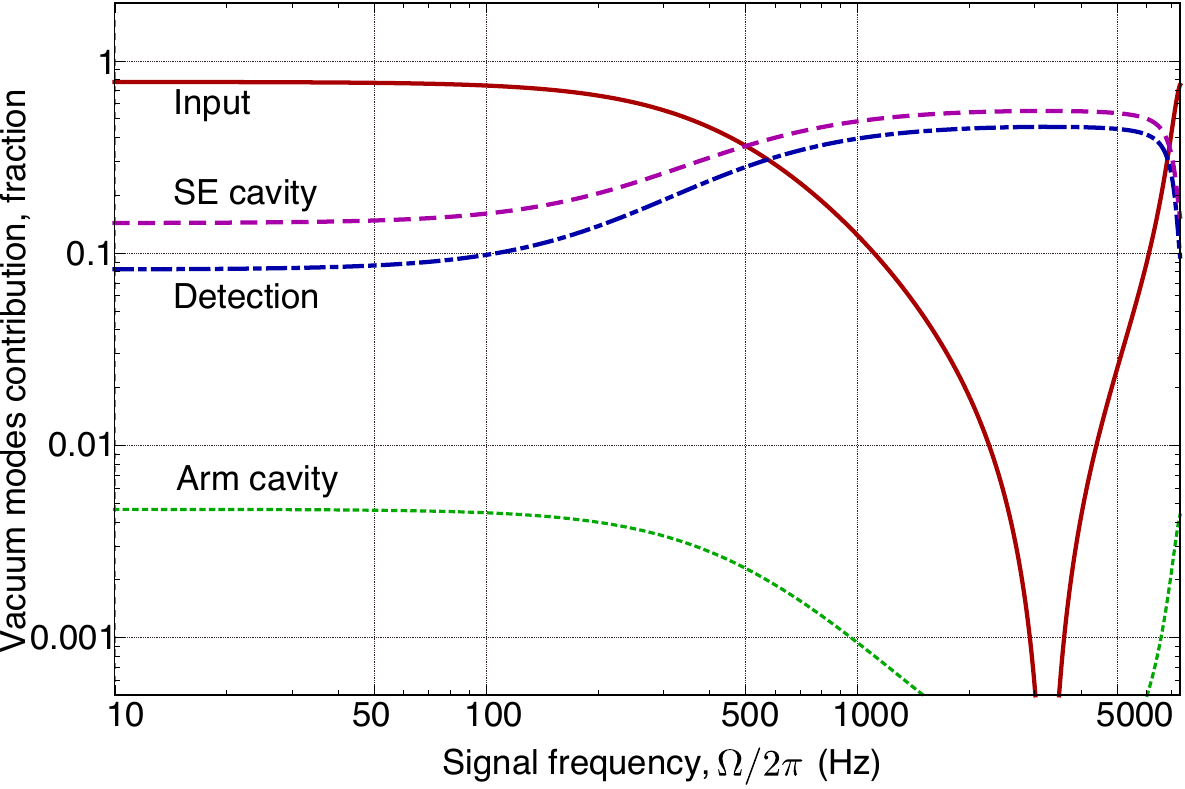}
            \caption{Relative contribution of different vacuum modes to the overall sensitivity of the detector at different frequencies.
            Input (solid red) vacuum mode defines the main sensitivity level, and the rest come from the various sources of loss: loss inside the SE cavity (dashed magenta), detection loss (dot-dashed blue) and arm cavity loss (dotted green).
            The parameters are taken according to Table S1: internal loss is 1500\,ppm single-trip, detection loss is 1\%, transmission of the end mirror is 100\,ppm (increased relative to Table S1 to emphasize the smallness of its influence on the sensitivity)}\label{fig:loss}
\end{figure}
As we discuss in the main text, quantum expander creates squeezing at high frequencies to counteract the effect of the detector's bandwidth.
When combined with external squeezing, quantum expander produces a high amount of squeezing at high frequencies, which imposes a strict requirements on reducing the optical losses.
The losses occur inside the detector: inside the arm cavity, and inside the SE cavity; as well as on the readout train: from the SE mirror to the detector.
The external squeezing additionally suffers from the injection loss.
On Fig.\ref{fig:loss} we show the contribution of different sources of loss as a function of frequency.
We note that the detection loss and loss inside the SE cavity are the most important contributions.
The detection loss currently is rather high, but the way to mitigate this loss by parametric amplification was proposed by Caves~\cite{Caves1981} and recently re-investigated experimentally~\cite{Knyazev2018}.
The idea of this approach is to amplify both the signal and the noise by the same amount before it experiences loss, such that the resulting noise is much above vacuum uncertainty, and the loss does not affect it significantly.
The simplest example of it is detecting some signal $G$ embedded in squeezed vacuum, with detection efficiency $\eta$:
\begin{equation}
  S = \eta (e^{-2r} + G^2) e^{2q} + 1 - \eta,
\end{equation}
where $r$ is the squeeze factor and $q$ is the Caves' amplification factor.
The SNR is given by $SNR = \eta e^{2q} \left(1 - \eta(1-e^{-2r}e^{2q})\right)^{-1}$.
Without amplification, $q=0$, in the limit of large squeezing $e^{-2r} \approx 0$ the SNR is limited to $SNR_{q=0}\leq\eta (1-\eta)^{-1}$.
%When the amplification is used, e.g. $q = r$, the SNR scales linearly with loss and amplification: $SNR_{q=r} = \eta e^{2r}$.
When the amplification is large, $q\rightarrow \infty$, the SNR becomes independent on the loss: $SNR_{q\rightarrow \infty} = e^{2r}$, and only benefits from initial squeezing.
The only source of detection loss that cannot be mitigated by Caves' amplification is the loss in the Faraday isolator used for injecting external squeezing.
We assume this to be a limitation in the detection loss, which corresponds to the 0.5\%\cite{Schreiber2017} mentioned in the main text.

Internal loss will be increased due to the additional optical surfaces of the nonlinear crystal and the absorption of the crystal.
While the actual contribution to the loss from such a crystal requires a separate investigation, we give an estimate based on the squeezing cavity design for the table-top experiments.
If the PPKTP crystal will be used, it's absorption is $\sim100$ ppm per cm depending on wavelength~\cite{Steinlechner2013}; the surfaces of the crystal will have to be coated with anti-reflecting coating to minimize the scattering loss.
We estimate that the current standard technology can bring this added loss on the level of 200--500\,ppm in single-pass.
\begin{figure}[t!]
        \includegraphics[width=1\linewidth]{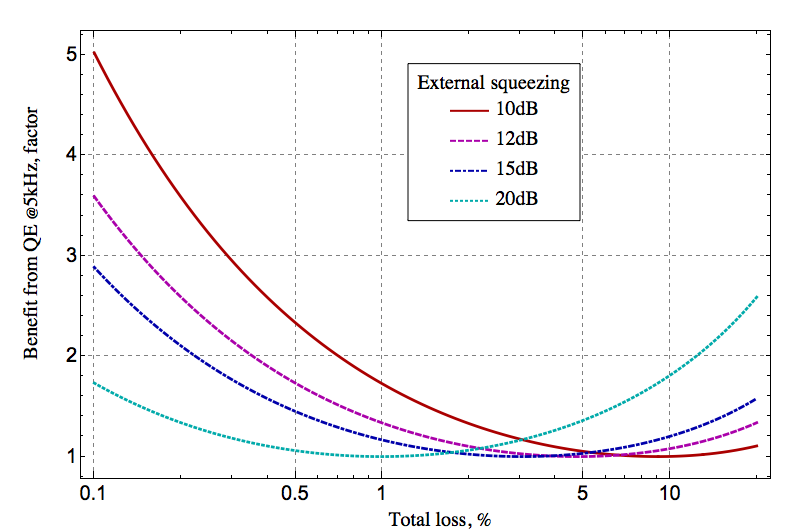}
            \caption{An improvement in the sensitivity of the detector by quantum expander, relative to the detector with external squeezing injection, depending on the amount of total loss (internal and readout). The higher is the external squeezing, the more stringent is the loss requirement for being able to benefit from using the quantum expander. The sensitivity depends in a non-trivial way on the losses, which is reflected in the benefit from QE shown on the figure.}\label{fig:benefit}
\end{figure}
\begin{figure}[t!]
        \includegraphics[width=1\linewidth]{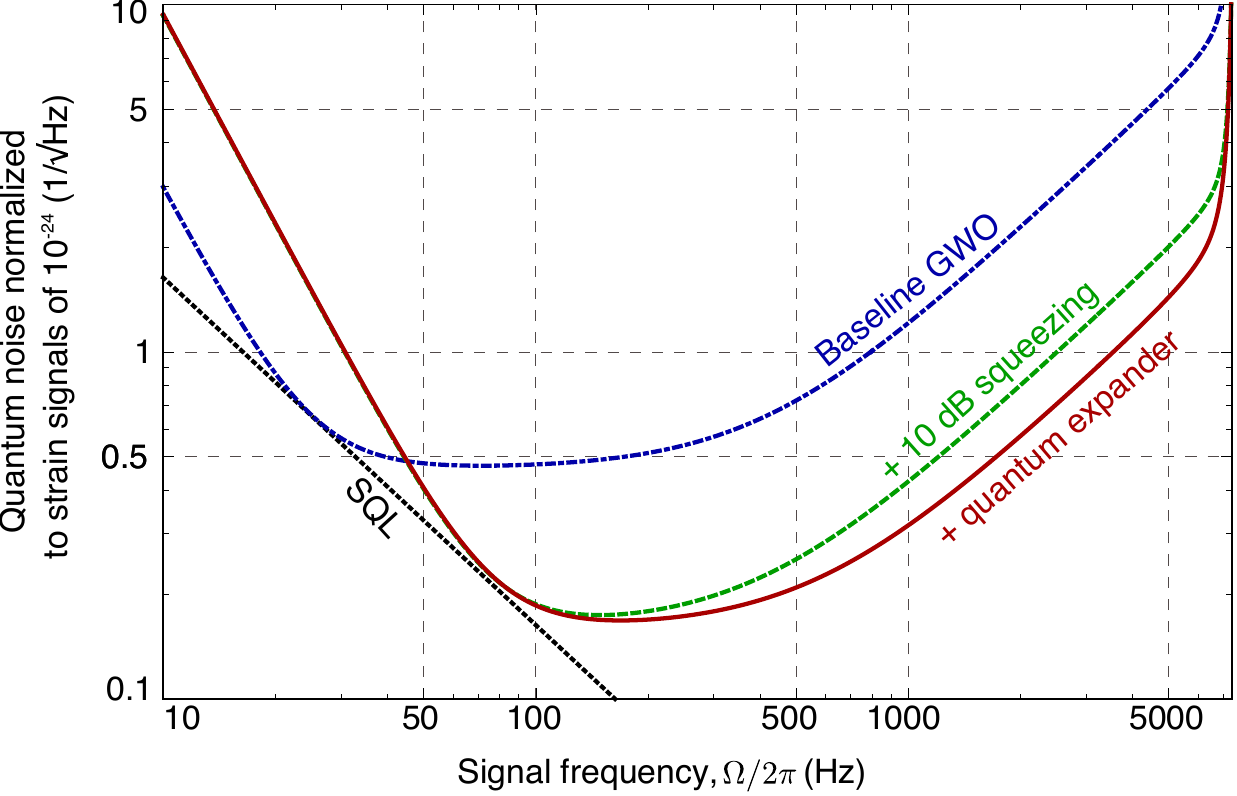}
            \caption{An example of sensitivity improvement in a particular design of a detector with 1\% of total loss and 10\,dB external squeezing injection, the parameters are given in Table S1.}\label{fig:benefit_example}
\end{figure}

We would like to emphasize, that not every configuration of the GWO will be able to get a significant benefit from quantum expansion when the external squeezing is in use.
Depending on the amount of loss, and amount of external squeezing injected, the benefit will vary.
The reason is an additional de-amplification of the signal in the quantum expander.
When the loss is high, the squeezing of the noise by quantum expander in addition to external squeezing might be not significant.
However, the parametric process inside the detector reduces the signal, hence the signal-to-noise ratio might even become reduced compared to the detector without quantum expander, if the sub-optimal parametric gain is chosen.
There always exists an optimal gain, for which the benefit is maximal.
If the loss is high, it might be optimal to amplify the signal (and anti-squeeze the noise), similar to the Caves' amplification discussed above.
We demonstrate possible improvements to the sensitivity in Fig. \ref{fig:benefit}.
We note, that this specific design is based on the benchmark parameters adopted by the LIGO-Virgo Collaboration, as presented in Table \ref{tab1}, and corresponds to the sensitivity as given in Fig. \ref{fig:benefit_example}.
In reality, the benefit from quantum expansion can be increased by optimizing the optical design (e.g. SE cavity length and mirrors' reflectivities).
The optimized sensitivity given by the quantum expander is a topic of future studies.
\begin{table}
  \begin{tabular}{ c | l | c | c}
    \hline
   \hline
     parameter & description & Baseline GWO & AdvLIGO \\
    \hline
    \hline
        $\lambda$ & optical wavelength & 1550nm & 1064nm \\ \hline
    $P_{\rm arm} = P_c/2$ & arm cavity light power & 4MW & 840kW \\ \hline
    $L_{\rm arm}$ & arm cavity length & 20 km & 4 km \\
    \hline
    $m$ & mirror mass & 200 kg & 40 kg \\
        \hline
    $L_{\rm SE}$ & SE cavity length & 56m & 56m \\
    \hline
    $T_i$ & input mirror power transmission & 0.07 & 0.014 \\
    \hline
     $T_{s} $& SE mirror power transmission & 0.35 & 0.35 \\
    \hline
    $T_{e} $& end mirror power transmission & 5ppm & 5ppm \\
    \hline
        $e^{2r}$ & external squeezing & 10\,dB & --- \\
    \hline
            $\lambda_s$ & loss inside SE cavity & 1500ppm & 1000ppm \\
    \hline
            $\eta$ & detection efficiency & 99\% & $\sim$85\% \\
    \hline
    \hline
  \end{tabular}
  \caption{ In order to plot the spectral densities in the paper we use the following set of parameters of some baseline GW observatory, without choosing a specific design from many possibilities of a 3-G topologies. We note that our double-cavity model uses effective parameters. In order to use this model for the Michelson topology, an effective light power inside the arm cavity has to be used: $P_c = 2 P_{\rm arm}$, where  $P_{\rm arm}$ is the power inside the arms of the Michelson topology\,\cite{Buonanno2003}}\label{tab1}
\end{table}

\subsection{Crystal inside the detector}
There are several issues to be taken into account with placing the crystal inside the SE cavity.

First, the size of crystal itself has to be large enough so that the optical beam does not clip on the edges of the crystal.
Currently the diameter of the beam inside the SE cavity is $\sim$\,2cm~\cite{Aasi2015}, with the focal point outside the SE cavity.
For comparison, as size of typical PPKTP crystal used in the squeezed light generation is 1$\times$2\,mm~\cite{Steinlechner2013}.
The crystal can be custom-made, or other nonlinear material can be used.
Further, the beam can be focused inside the SE crystal by changing the curvatures of the mirrors of SE cavity, without using additional optics.

Second, the absorption and scattering in the crystal are generally an important issue due to possible heating.
However, as in this design the detector operates at the dark port condition, there is no bright carrier field.

Third, the crystal has to be pumped with the frequency doubled parametric pump, which requires additional optical elements that would deliver the pump beam to the crystal and ensure the match between modes of the pump and the main beam.
This can be done in multiple ways.
As the wavelength of the pump is so different from the fundamental wavelength, it is possible to coat optical elements with different coatings, such that an additional cavity is formed by the SEM and ITM for the pump~\cite{Korobko2017b}.
Alternatively, the pump can be brought in by replacing the steering mirrors in the SE cavity with dichroic mirrors, transmissive for the frequency doubled pump.
In any case, no additional optics inside the main interferometer would be required.

In conclusion, while a non-linear crystal inside the interferometer is technologically challenging, we do not foresee fundamental problems, and expect our proposal for quantum expansion to motivate the future research and development work in this direction.

\section{Astrophysical analysis}
In this section, we give an illustrative example to estimate the capability of using the quantum expanders to detect the gravitational waves radiated by neutron star poster-merger remnants.
The method we used here follows the estimation procedure as described in~\cite{Yang2017,Miao2017}.
We perform a Monte Carlo simulation based on the following assumptions: first, the mass of each individual neutron star in a binary system follows an independent Gaussian distribution centered at $1.33 M_{\odot}$ with variance $0.09 M_{\odot}$.
The distributions of angular sky position, inclination and polarisation angles, and the initial phase of the source are assumed to be flat.
The searching range is assumed to be $1$\,Gpc and the event rate is taken to be $\approx 1\,{\rm Mpc}^{-3}{\rm Myr}^{-1}$.
Second, the post-merger waveform is assumed to be a parametrized damped oscillation, which depends on the equation of state of a neutron star, and in frequency domain it is given by the equation:
\begin{equation}
  h(f)=\frac{50{\rm Mpc}}{\pi d}h_p\frac{Q(2 f_p Q\cos\phi_0-(f_p-2ifQ)\sin\phi_0)}{f_p^2-4iff_pQ-4Q^2(f^2-f_p^2)},
\end{equation}
where $d$ is the source distance, $h_p$ is the peak value of the wave amplitude, $Q$ is the quality factor of the post-merger oscillation,$\phi_0, f_p$ are the initial phase and the peak frequency of the waveform, respectively.
Among them, $h_p, Q, f_p$ are parametrized by fitting with the results generated by numerical simulation~\cite{Bauswein2016} and they depend on the choice of equation of states. In the illustrative examples here, we make use of a relatively stiffer equation of state proposed in \cite{Shen1998}, where $Q=23.3, h_p\approx 5\times 10^{-22}$, and the peak frequency is given by:
\begin{equation}
  f_p=1{\rm kHz}\left(\frac{m_1+m_2}{M_{\odot}}\right)\left[a_2\left(\frac{R}{1{\rm km}}\right)^2+a_1\frac{R}{1{\rm km}}+a_0\right],
\end{equation}
where $R=14.42$\,km is the radius of each neutron star, and $m_{1,2}$ are their masses.
The parameters $a_2,a_1,a_0$ take the value of $5.503,-0.5495,0.0157$, respectively~\cite{Shen1998}.
We define the signal to noise ratio as:
\begin{equation}
{\rm SNR}=\int^{f_{\rm max}}_{f_{\rm min}}df\frac{|h(f)|^2}{S_{hh}(f)},
\end{equation}
where we take the integration range to be $f_{\rm min}=1000\,{\rm Hz}$, $f_{\rm max}=4000\,{\rm Hz}$.
We run 100 Monte-Carlo realizations each with $1000$ samples, corresponds to one-year observation.
We exclude the binaries with total mass larger than $3.45M_{\odot}$ since they will collapse into a black hole in a very short period of time, less than one period of post-merger oscillation.
For each different interferometer parameter set, we selected out the loudest event in each Monte-Carlo realization, set ${\rm SNR}=5$ as a threshold signal-to-noise ratio and produce the Figure 4 in the main text.

\section{Input-output relations}
In this section we derive the sensitivity based on the input-output formalism.
For simplicity in this section we ignore the effects of quantum radiation pressure noise and optical losses.
These will be included in the full transfer matrix description in Section S5.
Based on the obtained equations we give motivation for writing the Hamiltonian of the system in the Section S4.

Using the perturbation theory, we decompose the light field into a steady-state amplitude with amplitude $A_0$ and laser carrier frequency $\omega_0$ and a slowly varying noise amplitude $a(t)$ (see details in~\cite{Danilishin2012}):

\begin{eqnarray}
  &A(t)& = \sqrt{\frac{2\pi \hbar \omega_0}{\mathcal{A} c}} \left[ A_0 e^{-i\omega_0 t} + a(t) e^{-i\omega_0 t} \right] + {\rm h.c.}\\
  & \hat{a}(t)& = \int_{-\infty}^{\infty}  \hat{a}(\omega_0 + \Omega)e^{-i\Omega t} \frac{d \Omega}{2 \pi},
  \end{eqnarray}
  where $\mathcal{A}$ is the laser beam cross-section area, $\hbar$ is the reduced Plank constant.
It is helpful to consider the input-output relations of our system in the `two-photon formalism' \cite{Caves1985a, Schumaker1985a}, where the amplitude and phase quadrature amplitudes $\hat{a}^{(c)}$ and $\hat{a}^{(s)}$ of the modulation field at frequency $\Omega$ are linked to the optical fields $\hat{a}(\omega \pm \Omega)$ via
\begin{eqnarray}
\hat{a}^{(c)}(\Omega) &=& \frac{\hat{a}(\omega +  \Omega) + \hat{a}^\dag(\omega - \Omega))}{\sqrt{2}}\, , \\
\hat{a}^{(s)}(\Omega) &=& \frac{ \hat{a}(\omega + \Omega) - \hat{a}^\dag(\omega - \Omega)}{i\sqrt{2}}\, .
\end{eqnarray}

These operators obey the commutation relation
  \begin{eqnarray}
  &[a_x(\Omega), a_x(\Omega')]& = [a_y(\Omega), a_y(\Omega')] = 0,\\
   &[a_x(\Omega), a_y(\Omega')]& = [a_x(\Omega), a_y(\Omega')] = 2 \pi i \delta(\Omega + \Omega').
  \end{eqnarray}

We make several simplifications to the notation: as we are primarily interested in the phase quadrature, we will omit index $(s)$ in equations below; we also omit the hats on the operators for brevity, although all the fields are quantised; we consider only the noise fields in the frequency domain, so we don't write that in the equations explicitly: \textit{e.g.} $\hat{a}^{\rm (s)}(\Omega) \rightarrow a$.

The signal we consider is a phase modulation on the light field induced by motion of the mirror with infinite mass caused by an external force.
This modulation adds a phase shift on the light reflected off the movable mirror: $E_{\rm refl} = E_{\rm in}e^{2 i k x(\Omega)}\approx E_{\rm in}(1 + 2 i k_p x(\Omega))$, where $k_p$ is the light's wave vector, $E_{\rm refl, in}$ are the amplitudes of the reflected and incident light fields, and $x(\Omega)$ is a small mirror displacement.
The signal appears only in the equations for the phase quadrature of the light field.

We model the parametric amplification process as a simple linear amplification of amplitude quadrature of the light by some factor $e^q$, without considering the effects of the parametric pump and the finite size of a crystal. In the full model in section 5 we also will introduce the possibility to tune the amplification quadrature.
\begin{figure}[t!]
        \includegraphics{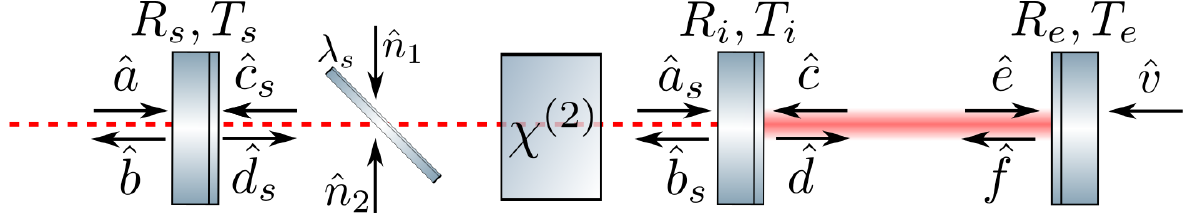}
            \caption{Quantum fields in the model of a two-cavity system. $R_{s,i,e},T_{s,i,e}$ are the amplitude reflectivities and transmissivities of the signal extraction, input and end test mirrors correspondingly; a beam-splitter with power reflectivity $\lambda_s$ represents a source of intra-cavity loss, which causes vacuum noises $\hat{n}_{1,2}$ to enter the system.}\label{fig:setup_theory}
\end{figure}
With this in mind we start with writing down the steady-state input-output relations~\cite{Caves1985a, Schumaker1985a} for the quantum fluctuations of the phase quadrature of the light field, for the cavity cavity model depicted of Fig.~\ref{fig:setup_theory}.
For the detailed explanation of the approach we refer the reader to the review by Danilishin and Khalili \cite{Danilishin2012}.
We choose the arm cavity to be tuned on resonance, so that for $\Omega = 0$ it has the maximal light power inside.

\begin{eqnarray}
&d_s& = T_s a + R_s c_s,\\
&a_s& = d_s e^{-q} e^{i\varphi} e^{i\Omega \tau_{\rm SE}},\\
&b_s& = T_i c +R_i a_s,\\
&c& = d e^{2i \Omega \tau_{\rm arm}} + 2i k_p E x e^{i \Omega \tau_{\rm arm}},\\
&b_s& = -R_i a_s + T_i c,\\
&c_s& = b_s e^{-q} e^{i\varphi} e^{i\Omega \tau_{\rm SE}},\\
&b& = -R_s a + T_s c_s,
\end{eqnarray}
where $R_{i,s} = \sqrt{R_{\rm ITM, SE}}, T_{i,s}=\sqrt{T_{\rm ITM, SE}}$ are the amplitude reflectivity and transmissivity of input test mirror and signal-extraction mirror; $q$ is an amplification factor on the single pass through the crystal; $\tau_{\rm arm, SE} = L_{\rm arm, SE}/c$ is the single trip time in arm cavity of length $L_{\rm arm}$ and signal extraction cavity of length $L_{\rm SE}$, with $c$ being the speed of light; $\varphi=\pi/2$ is the tuning of the SE cavity with respect to the arm cavity;  $x$ is a small displacement of the end mirror due to the GW signal, $E$ is the large classical amplitude of field inside the arm cavity and $k_p$ is the wave vector of the carrier light field.

We find a solution to these equation, splitting the output $b$ into the noise part $b_n$ and signal $G_{\rm out}$: $b = b_n + X_{\rm out}$.
\begin{eqnarray}\label{eq:solgen}
&b_n& = \mathcal{R}_a(\Omega) a(\Omega)= -\frac{e^{2i\varphi}e^{2i\Omega \tau_{\rm SE}} (e^{2i\Omega \tau_{\rm arm}} - R_i) + e^{2q} (e^{2 i\Omega \tau_{\rm arm}} R_i - 1)}{e^{2q} (e^{2i\Omega \tau_{\rm arm}} R_i - 1) + e^{2i\varphi}e^{2i\Omega \tau_{\rm SE}} (e^{2i\Omega \tau_{\rm arm}} - R_i) R_s} a(\Omega),\\
&X_{\rm out}& = \mathcal{T}(\Omega) x(\Omega) = \frac{2 i k_p E e^{i\varphi} e^{i\Omega \tau_{\rm SE}} e^{i\Omega \tau_{\rm arm}} e^{q} T_i T_s}{e^{2q} (e^{2i\Omega \tau_{\rm arm}} R_i - 1) + e^{2i\varphi}e^{2i\Omega \tau_{\rm SE}} (e^{2i\Omega \tau_{\rm arm}} - R_i) R_s} x(\Omega),
\end{eqnarray}
where $\mathcal{R}_a(\Omega), \mathcal{T}(\Omega)$ are the noise and signal optical transfer functions correspondingly.

We can obtain an intuitive expression for these functions by doing several approximations.
We assume $\Omega \tau_{\rm arm} \ll 1$, $\Omega \tau_{\rm SE} \ll 1$, so $e^{i\Omega \tau_{\rm arm, SE}} \approx 1+ i\Omega \tau_{\rm arm, SE}$; and $T_{i,s} \ll 1$, so $R_{i} \approx 1-T_{i}^2/2 = 1 - 2 \gamma_{\rm arm} \tau_{\rm arm}$, $R_{s} \approx 1-T_{s}^2/2 = 1 - 2 \gamma \tau_{\rm arm}$, where $\gamma_{\rm arm},\gamma$ are the arm cavity and the signal-extraction cavity linewidth, respectively; a single-pass optical gain is small: $q\ll 1$, so $e^{q}\approx 1+q = 1 + \chi \tau_{\rm SE}$, where $\chi$ is an effective parametric gain.

With these approximations equations (18-19) can be simplified to
\begin{eqnarray}
&\mathcal{R}_a(\Omega)& = \frac{(\gamma - \chi)\Omega + i(\Omega^2-\omega_s^2)}{(\gamma + \chi )\Omega - i(\Omega^2-\omega_s^2)}\\
&\mathcal{T}(\Omega)& = -\frac{4ik_pE}{\sqrt{\tau_{\rm arm}}}\frac{\sqrt{\gamma}\omega_s}{(\gamma + \chi)\Omega - i(\Omega^2-\omega_s^2)},
\end{eqnarray}
where we defined a sloshing frequency $\omega_s = c\sqrt{T^2_i/(4 L_{\rm SE} L_{\rm arm})}$.
Notice that these equations correspond to Eq. (6) in the main text.
This helps us to construct a Hamiltonian in the next Section, which would correspond to this model system.

We would like to point out the limits of this approximation: it is valid only until sloshing and signal frequencies are much smaller than the free spectral range of the arm cavity: $\Omega, \omega_s \ll c/{L_{\rm arm}}$.
This condition sets a limit on the transmissivity of the ITM: $T_i^2\ll L_{\rm SE}/L_{\rm arm}$.
This restricts the applicability of the derived equations to a detector with a relatively short arm length (e.g. AdvancedLIGO), while a longer detector (as baseline GWO considered in Table S1) would require a more sophisticated expression with the higher FSR of the arm cavity taken into account.
The assumption of a small transmission of the SE mirror is often not valid in real designs, which would lead to additional contributions in the noise spectrum.
We perform the full analysis in Section 5.

\section{Hamiltonian approach}
In this section we derive the sensitivity of the detector (Eq. 6 of the main text) from the Hamiltonian of the system.
The Hamiltonian is based on the input-output formalism, derived in the previous section, where a set of approximations was made.
These approximations restrict the analysis to the case when only two modes are taken into account: one in the arm cavity and one in the signal extraction cavity.

\begin{eqnarray}
&\hat{H}& = \hat{H}_0 + \hat{H}_{\rm int} + \hat{H}_\gamma + \hat{H}_{x} - F_{\rm GW} x; \\
&\hat{H}_0& = \hbar \omega_0 \hat{a}^\dag \hat{a} + \hbar \omega_0 \hat{a}_q^\dag \hat{a}_q;\\
&\hat{H}_{\rm int}& =  \hbar \omega_s \hat{a}_q^\dag \hat{a} + \frac{1}{2}\hbar \kappa \beta e^{-2i\omega_0 t} \hat{a}_q^\dag \hat{a}_q^\dag e^{i\phi} + h.c.;\\
&\hat{H}_\gamma& = i \hbar \sqrt{2\gamma} \int_{-\infty}^{\infty} \left(\hat{a}_q^\dag(\omega)\hat{a}_{\rm in}(\omega) - \hat{a}_{\rm in}^\dag(\omega)\hat{a}_q(\omega)\right)d\omega;\\
&\hat{H}_x& = - \hat{F}_{\rm rp} \hat{x} = -\hbar G_0 \hat{a}^\dag \hat{a} \hat{x},
\end{eqnarray}
where $\hat{a}, \hat{a}_q$ are the arm cavity and SE cavity modes, and $\omega_0$ is their natural resonance frequency; $\omega_s = c\sqrt{T_{\rm ITM}/(4 L_{\rm SE} L_{\rm arm})}$ is the coupling rate between two cavities, $T_{\rm ITM}$ is the transmission of the front mirror of the arm cavity, $L_{\rm SE}, L_{\rm arm}$ are the lengths of the signal extraction and arm cavity, respectively;  $\gamma = c T_{\rm SE}/(4 L_{\rm SE})$ is the coupling rate of the SE mode to the continuum of input modes $\hat{a}_{\rm in}$; $x$ is the displacement of the test mass partially in reaction to the gravitational-wave tidal force $F_{\rm GW}$; the mirror motion $x$ is coupled via the radiation-pressure force $\hat{F}_{\rm rp}$ to the cavity mode with strength $G_0=\omega_0/L_{\rm arm}$; $\kappa$ is the coupling strength due to a crystal nonlinearity under a second harmonic pump field with amplitude $\beta$ and phase $\phi$.
The pump field is assumed to be classical and its depletion is neglected.
The effect of the back-action noise can be neglected, so displacement of the mirror is coupled only to a GW strain: $x=h_0/L_{\rm arm}$.

We obtain the Langevin equations of motion for the cavity modes in the frame rotating at $\omega_0$ and expand the quantum amplitudes into a sum of large classical amplitude and small quantum fluctuation, $\hat{a} \rightarrow A + \hat{a}$:
\begin{eqnarray}
  &\dot{\hat{a}}& = -i\omega_s \hat{a}_q + i  G h_0;\\
  &\dot{\hat{a}}_q& = -i\omega_s\hat{a} -  \gamma \hat{a}_q +  \sqrt{2 \gamma} a_{\rm in} - i \chi \hat{a}_s^\dag e^{i\phi};\\
  &\hat{a}_{\rm out}& = -\hat{a}_{\rm in} + \sqrt{2\gamma} \hat{a}_q.
\end{eqnarray}
where we defined an effective coupling strength of GW signal strain $G = \sqrt{2 P_c L_{\rm arm}\omega_0/(\hbar c)}$ and optical power inside the arm cavity $P_c = \hbar \omega_0 \bar{a}$, with $\bar{a}$ being an average amplitude of the mode $\hat{a}$; and the effective parametric gain $\chi = \kappa \beta$,

As we are interested in the spectral properties of the system, we transform into a Fourier domain: $\dot{\hat{a}}(t)\rightarrow -i\Omega \hat{a}(\Omega)$.
The outgoing light is measured by a homodyne detector, which measures the quadratures of the light, that are defined as:
\begin{equation}
  \hat{a}^{(c)} = \frac{\hat{a}(\Omega) + \hat{a}^{\dag}(-\Omega)}{\sqrt{2}}, \qquad \hat{a}^{(s)} = \frac{\hat{a}(\Omega) - \hat{a}^{\dag}(-\Omega)}{i\sqrt{2}}
\end{equation}
 We obtain the input-output relations for the two quadratures by solving Eqs.(27-29):
\begin{eqnarray}
  \hat{a}_{\rm out}^{(c)}(\Omega) &=& \hat{a}_{\rm in}^{(c)}(\Omega)\frac{(\gamma - \chi + i\Omega)\Omega - i\omega_s^2}{(\gamma + \chi - i\Omega)\Omega + i\omega_s^2} + h_0(\Omega) \frac{2iG\sqrt{\gamma}\omega_s}{(\gamma + \chi - i\Omega)\Omega + i\omega_s^2}\\
  &=& \hat{a}_{\rm in}^{(c)}(\Omega) \mathcal{R}_a(\Omega) + h_0(\Omega)\mathcal{T}(\Omega)\\
  \hat{a}_{\rm out}^{(s)}(\Omega) &=& \hat{a}_{\rm in}^{(s)}(\Omega)\frac{(\gamma + \chi + i\Omega)\Omega - i\omega_s^2}{(\gamma - \chi - i\Omega)\Omega + i\omega_s^2}.
\end{eqnarray}

From these input-output relations we can obtain the sensitivity, by computing the spectral densities.
We define the spectral density of the field $\hat{a}(\Omega)$ as:
\begin{equation}
S_{a} (\Omega) \delta(\Omega - \Omega') = \frac{1}{2}\left<\hat{a}(\Omega)\hat{a}(\Omega') + \hat{a}(\Omega')\hat{a}(\Omega)\right>,
\end{equation}
Then the spectral density the output noise $\hat{a}_{\rm out}^{(c)}(\Omega)$ is:
\begin{equation}
S_{\rm out} (\Omega) = S_{\rm in} (\Omega) |\mathcal{R}_a(\Omega)|^2,
\end{equation}
where $S_{\rm in} (\Omega)$ is the spectral density of incoming light field, which we assume here to be vacuum: $S_{\rm in} (\Omega) = 1$.
Assuming that we squeeze the signal quadrature of the light: $\phi = -\pi/2$, we obtain the following noise spectral density
\begin{equation}
S_{\rm out} (\Omega) = 1 - \frac{4 \gamma \chi \Omega^2}{(\gamma + \chi)^2 \Omega^2 + (\Omega^2 - \omega_s^2)^2}
\end{equation}
and signal transfer function:
\begin{equation}
  |\mathcal{T}(\Omega)|^2 = \frac{4 G^2 \gamma \omega_s^2}{(\gamma + \chi)^2 \Omega^2 + (\Omega^2 - \omega_s^2)^2}.
\end{equation}
The total strain sensitivity is given by the noise normalised to the signal transfer function:
\begin{equation}
  S_h(\Omega) = \frac{(\gamma-\chi)^2\Omega^2 + (\Omega^2-\omega_s^2)^2}{4 G^2 \gamma \omega_s^2},
\end{equation}
which is the equation (9) in the Main text.

\section{Transfer matrix approach to full description}
In this section we use the transfer matrix approach~\cite{Danilishin2012} to compute the sensitivity of the detector taking into account the radiation pressure noise and optical losses.
We start from the same point as in the section S3, but write down the input-output relations as propagation of the field amplitudes in terms of transfer matrices for each optical element. The description is broader than strictly needed to compute the spectral density in the main text (e.g. it includes the effects of dynamical back action), but we find it helpful to use a general approach.

\subsection{Input-output relations}
We describe a two-cavity system, as shown on Fig.\ref{fig:setup_theory}, in terms of input and output quantum fields.
Based on two-photon quadrature amplitudes we define the vector ${\hat{\mathrm a}}(\Omega) = \{\hat{a}^{(c)}(\Omega), \hat{a}^{(s)}(\Omega)\}^{\rm T}$. The signal extraction cavity can rotate the quadratures due to it's detuning from resonance. The optical parametric amplification process also squeezes and rotates the quadratures.
The effect of the signal recycling cavity can be described as a set of rotations and squeezing operations:
\begin{eqnarray}
  {\hat{\mathrm a}}_s &=& O(\varphi)O(\theta)\mathcal{S}O^\dag(\theta)O(\phi) (\sqrt{1-\lambda_s} {\hat{d}}_s + \sqrt{\lambda_s} {\hat{n}}_1) e^{i\Omega \tau_{\rm SE}}\, ,\\
  \mathrm{\hat{b}}_s &=& -R_i \mathrm{\hat{a}}_s + T_i \mathrm{\hat{c}},\\
  \mathrm{\hat{c}}_s  &=&\sqrt{1-\lambda_s} O(\phi)O(\theta)\mathcal{S}O^\dag(\theta)O(\varphi) \mathrm{\hat{b}}_s e^{i\Omega \tau_{\rm SE}} + \sqrt{\lambda_s} \mathrm{\hat{n}}_2\, ,\\
  \mathrm{\hat{d}}_s &=& T_s \mathrm{\hat{a}} + R_s \mathrm{\hat{c}}_s,
\end{eqnarray}
where we denote the amplitude reflectivity and transmissivity of the signal recycling and input mirrors by $R_{s, i}, T_{s,i }$, the power loss inside the cavity (before the crystal) is $\lambda_s$; signal recycling cavity global delay $\tau_{\rm SE} = L_{\rm SE}/c$ and the phase delay due to the cavity detuning before and after the crystal by $\phi, \varphi$.
We now introduce the squeeze angle $\theta$ and the rotation matrix
\begin{equation}
\forall \phi, \quad O(\phi) = \{\{\cos \phi, -\sin \phi\},\{\sin \phi, \cos \phi\}\}
\end{equation}
\begin{equation}
\mathcal{Y} = O(\pi/2) = \{\{0, -1\},\{1, 0\}\}
\end{equation}
and squeezing matrix
\begin{equation}
\mathcal{S} = \{\{e^q, 0\},\{0, e^{-q}\}\},
\end{equation}
with $q$ being the single-pass squeeze factor.

For the arm cavity the corresponding set of equations reads
\begin{eqnarray}
  \mathrm{\hat{b}}  &=& -R_s \mathrm{\hat{a}} + T_s \mathrm{\hat{c}}_s\, ,\\
  \mathrm{\hat{d}}  &=& R_i \mathrm{\hat{c}} + T_i \mathrm{\hat{a}}\, ,\\
  \mathrm{\hat{c}} &=& O(\delta_{\rm arm} \tau_{\rm arm})\mathrm{\hat{f}} e^{i\Omega \tau_{\rm arm}}\, ,\\
  \mathrm{\hat{e}} &=& O(\delta_{\rm arm} \tau_{\rm arm})\mathrm{\hat{d}}e^{i\Omega \tau_{\rm arm}}\, ,\\
  \mathrm{\hat{f}} &=& R_e\mathrm{\hat{e}} + T_e \mathrm{\hat{v}}   + 2 k R_e O(\pi/2)\mathrm{E} \hat{x}_-(\Omega)\, , %
\end{eqnarray}
where $k = \omega/c$ is the wave vector of the main field, $\delta_{\rm arm}$ is the arm cavity detuning  and $\tau_{\rm arm} = L_{\rm arm}/c$  is the propagation time with $L_{\rm arm}$ being the length of the arm cavity, and $c$ the speed of light.
The field $\mathrm{E}$ corresponds to the classical amplitude of the field impinging on the end mirror.
%This set of equations can be resolved for the outgoing field $\mathrm{\hat{b}}$ and intra-cavity fields $\mathrm{\hat{c}},\mathrm{\hat{d}},\mathrm{\hat{e}},\mathrm{\hat{f}}$.

We find the solution to these equations, first for the complex transmissivity and reflectivity of the signal recycling cavity
\begin{eqnarray}
  \mathrm{\hat{b}}_s &=& \mathcal{D}_b \left[ -R_i T_s\mathcal{M}[\varphi, \phi] \mathrm{\hat{a}} + T_i \mathrm{\hat{c}}\right]\, ,\\
  \mathrm{\hat{d}}_s &=& \mathcal{D}_d \left[ R_s T_i\mathcal{M}[\phi, \varphi] \mathrm{\hat{c}} + T_s \mathrm{\hat{a}}\right]\, ,\\
  \mathrm{\hat{a}}_s &=& \mathcal{M}[\varphi,\phi]\mathcal{D}_d \left[ R_s T_i\mathcal{M}[\phi, \varphi] \mathrm{\hat{c}} + T_s \mathrm{\hat{a}}\right]\, ,\\
  \mathrm{\hat{c}}_s &=& \mathcal{M}[\phi,\varphi]\mathcal{D}_b \left[ -R_i T_s\mathcal{M}[\varphi, \phi] \mathrm{\hat{a}} + T_i \mathrm{\hat{c}}\right]\, ,\\
\end{eqnarray}
where we defined
\begin{eqnarray}
  \mathcal{M}[\phi,\psi] &=& O(\phi)O(\theta)\mathcal{S}O^\dag(\theta)O(\psi)e^{i\Omega \tau_{\rm SE}}, \forall \phi, \psi\, ,\\
  \mathcal{D}_b &=& \left( \mathcal{I} + R_i R_s (1-\lambda_s)  \mathcal{M}[\varphi,\phi]\mathcal{M}[\phi,\varphi]\right)^{-1}\, ,\\
  \mathcal{D}_d &=& \left( \mathcal{I} + R_i R_s  (1-\lambda_s) \mathcal{M}[\phi,\varphi]\mathcal{M}[\varphi,\phi]\right)^{-1}\, .
\end{eqnarray}
That provides the input-output relations for the signal extraction cavity
\begin{eqnarray}
  \mathrm{\hat{b}} &=& -\mathcal{R}_b \mathrm{\hat{a}} + \mathcal{T}_b \mathrm{\hat{c}} + \mathcal{L}_{b1}\mathrm{\hat{n}}_1 + \mathcal{L}_{b2}\mathrm{\hat{n}}_2\, ,\\
  \mathrm{\hat{d}} &=& \mathcal{R}_d \mathrm{\hat{c}} + \mathcal{T}_d \mathrm{\hat{a}}+ \mathcal{L}_{d1}\mathrm{\hat{n}}_1 + \mathcal{L}_{d2}\mathrm{\hat{n}}_2\, ,
\end{eqnarray}
where we introduced the transfer matrices for the fields
\begin{eqnarray}
  \mathcal{R}_b &=& R_s + R_i T_s^2  (1-\lambda_s) \mathcal{M}[\phi,\varphi]\mathcal{D}_b \mathcal{M}[\varphi,\phi]\, ,\\
  \mathcal{R}_d &=& R_i + R_s T_i^2 (1-\lambda_s) \mathcal{M}[\varphi,\phi]\mathcal{D}_d \mathcal{M}[\phi,\varphi]\, ,\\
  \mathcal{T}_b &=& T_i T_s \sqrt{1-\lambda_s} \mathcal{M}[\phi,\varphi]\mathcal{D}_b\, ,\\
  \mathcal{T}_d &=& T_i T_s \sqrt{1-\lambda_s} \mathcal{M}[\varphi,\phi]\mathcal{D}_d\,,\\
  \mathcal{L}_{d1} &=& - T_i R_i R_s \sqrt{1-\lambda_s} \lambda_s \mathcal{M}[\phi,\varphi]\mathcal{D}_d \mathcal{M}[\varphi,\phi] + \sqrt{\lambda_s}\, ,\\
  \mathcal{L}_{d2} &=& T_i R_s \sqrt{\lambda_s(1-\lambda_s)} \lambda_s \mathcal{M}[\phi,\varphi]\mathcal{D}_d.
\end{eqnarray}
Now we can derive the fields for the arm cavity yielding
\begin{eqnarray}
\mathrm{\hat{c}} &=& R_e \mathcal{D}_c O(\delta_{\rm arm} \tau_{\rm arm})^2 \mathcal{T}_d \mathrm{\hat{a}} e^{2 i \Omega \tau_{\rm arm}} +  T_e \mathcal{D}_cO(\delta_{\rm arm} \tau_{\rm arm})\mathrm{\hat{v}} e^{ i \Omega \tau_{\rm arm}} + \nonumber\\ &+& R_e \mathcal{D}_c O(\delta_{\rm arm} \tau_{\rm arm})^2 \left(\mathcal{L}_{d1} \mathrm{\hat{n}_1} + \mathcal{L}_{d2} \mathrm{\hat{n}_2}  \right) e^{2 i \Omega \tau_{\rm arm}} + \nonumber\\&+& 2 k R_e \mathcal{D}_c O(\delta_{\rm arm} \tau_{\rm arm}) \mathcal{Y}\mathrm{E} \hat{x}_-(\Omega) e^{i\Omega\tau_{\rm arm}}\\
\mathrm{\hat{e}} &=& \mathcal{D}_e O(\delta_{\rm arm} \tau_{\rm arm}) \mathcal{T}_d \mathrm{\hat{a}} e^{i \Omega \tau_{\rm arm}} + T_e \mathcal{D}_e O(\delta_{\rm arm} \tau_{\rm arm}) \mathcal{R}_d O(\delta_{\rm arm} \tau_{\rm arm}) \mathrm{\hat{v}} e^{2 i \Omega \tau_{\rm arm}} + \nonumber \\ &+&\mathcal{D}_e O(\delta_{\rm arm} \tau_{\rm arm}) \left(\mathcal{L}_{d1} \mathrm{\hat{n}_1} + \mathcal{L}_{d2} \mathrm{\hat{n}_2}  \right) e^{i \Omega \tau_{\rm arm}}+ \nonumber\\ &+&2 k R_e \mathcal{D}_e O(\delta_{\rm arm} \tau_{\rm arm}) \mathcal{R}_d O(\delta_{\rm arm} \tau_{\rm arm}) \mathcal{Y}\mathrm{E} \hat{x}_-(\Omega) e^{2i\Omega\tau_{\rm arm}}
\end{eqnarray}
where
\begin{eqnarray}
\mathcal{D}_c &=&  \left( \mathcal{I} - R_eO(\delta_{\rm arm} \tau_{\rm arm})^2 \mathcal{R}_d e^{2i\Omega \tau_{\rm arm}}\right)^{-1}\\
\mathcal{D}_e &=&  \left( \mathcal{I} - R_eO(\delta_{\rm arm} \tau_{\rm arm}) \mathcal{R}_d O(\delta_{\rm arm} \tau_{\rm arm}) e^{2i\Omega \tau_{\rm arm}}\right)^{-1}.
\end{eqnarray}

Finally, we find the outgoing field to be
\begin{equation}
\mathrm{\hat{b}} = -\mathcal{R} \mathrm{\hat{a}}  + \mathcal{T}\mathrm{\hat{v}} + \mathcal{Z} \hat{x}_-(\Omega) + \mathcal{L}_{b1} \mathrm{\hat{n}_1} + \mathcal{L}_{b2} \mathrm{\hat{n}_2}
\end{equation}
where we defined the transfer matrices:
\begin{eqnarray}
\mathcal{R} &=&  \mathcal{R}_b - R_e\mathcal{T}_b \mathcal{D}_cO(\delta_{\rm arm} \tau_{\rm arm})^2 \mathcal{T}_d e^{2i\Omega\tau_{\rm arm}}\, ,\\
\mathcal{T} &=&  T_e \mathcal{T}_b\mathcal{D}_cO(\delta_{\rm arm} \tau_{\rm arm})e^{ i \Omega \tau_{\rm arm}}\, ,\\
\mathcal{Z} &=&  2 k R_e \mathcal{T}_b\mathcal{D}_c O(\delta_{\rm arm} \tau_{\rm arm}) \mathcal{Y}\mathrm{E} e^{i\Omega\tau_{\rm arm}}\, ,\\
\mathcal{L}_{b1} &=&  -T_s R_i \sqrt{1-\lambda_s} \lambda_s \mathcal{M}[\varphi,\phi]\mathcal{D}_b \mathcal{M}[\phi,\varphi]\, ,\\
\mathcal{L}_{b2} &=&  T_s R_i R_s \sqrt{1-\lambda_s} \lambda_s \mathcal{M}[\varphi,\phi]\mathcal{D}_b \mathcal{M}[\phi,\varphi] - \sqrt{\lambda_s}\,.
\end{eqnarray}

\subsection{Radiation pressure}
The radiation pressure force acting on the mirrors has three contributions. First, there is a constant force due to the classical high-power optical field. It induces a constant shift of the mirror, which can be compensated with classical feedback.
Second, there is a dynamical classical part, which is amplified by opto-mechanical parametric amplification and which belongs to the optical spring, and third a fluctuating force due to the uncertainty in the amplitude quadrature of the light.
The latter corresponds to the quantum back-action force of the carrier light. Following \,\cite{Buonanno2003}, we assume the input test mass to be fixed, and twice the back action imposed on the back mirror instead (which leads to introduction of effective light power). Such approximation is valid when the transmission of front mirror is small, such that the amplitudes of the fields acting on the front and back mirrors are almost equal (which is the case in our consideration).

\begin{equation}
F^{ba} = \hbar k (\mathrm{E}^\dag \mathrm{\hat{e}}(\Omega) + \mathrm{F}^\dag \mathrm{\hat{f}}(\Omega)) = F_{fl}(\Omega) - \mathcal{K}(\Omega) x_-(\Omega)\,.
\end{equation}
where we split the back-action into the noise part $F_{fl}(\Omega)$ and position-dependent optical spring force with spring constant $\mathcal{K}(\Omega)$.
Taking into account that $\mathrm{F}  =  R_e \mathrm{E}$, we find the equations for these contributions:

\begin{eqnarray}
  F^{fl}(\Omega) &=& \hbar k (1+R_e^2) \mathrm{E}^\dag \mathcal{D}_e O(\delta_{\rm arm} \tau_{\rm arm}) e^{i \Omega \tau_{\rm arm}} \left(\mathcal{T}_d \mathrm{\hat{a}} + \mathcal{L}_{d1} \mathrm{\hat{n}_1} + \mathcal{L}_{d2} \mathrm{\hat{n}_2}  \right)  + \nonumber \\ &+&  \hbar k T_e \mathrm{E}^\dag \mathcal{L}_v \mathrm{\hat{v}};\\
  \mathcal{L}_v  &=&  (1+R_e^2) \mathcal{D}_e O(\delta_{\rm arm} \tau_{\rm arm}) \mathcal{R}_d O(\delta_{\rm arm} \tau_{\rm arm}) e^{i \Omega \tau_{\rm arm}} +R_e;\\
  \mathcal{K}(\Omega) &=& -2\hbar k^2 (1+R_e^2) R_e \mathrm{E}^\dag \mathcal{D}_e O(\delta_{\rm arm} \tau_{\rm arm}) \mathcal{R}_d O(\delta_{\rm arm} \tau_{\rm arm}) \mathcal{Y}\mathrm{E} e^{2i\Omega\tau_{\rm arm}} - \nonumber\\ &-& 2 \hbar k^2 R_e^2 \mathrm{E}^\dag \mathcal{Y} \mathrm{E}.
\end{eqnarray}
Without loss of generality we choose the phase of the classical amplitude such that:
\begin{equation}
\mathrm{E} = \sqrt{2}E \{1, 0\}^{\mathrm T}
\end{equation}
where the amplitude $E$ is connected to the power in the cavity as $P_c = 2 P_{\rm arm} = \hbar \omega_p |E|^2$, where $P_{\rm arm}$ is a power in the corresponding Michelson interferometer\,\cite{Buonanno2003}.

The equation of motion for the test mass taking into account the radiation pressure force:
\begin{equation}
\hat{x}_-(\Omega) = \chi(\Omega) \left[ F^{fl}(\Omega) - \mathcal{K}(\Omega) x_-(\Omega) \right]\, ,
\end{equation}
which allows us to introduce an effective susceptibility:
\begin{equation}
\chi_{\rm eff}(\Omega) = (\chi^{-1} + \mathcal{K}(\Omega))^{-1}\, ,
\end{equation}
such that $x_-(\Omega) = \chi_{\rm eff}(\Omega) F^{fl}(\Omega)$.

\subsection{Detection}
The presence of optical loss in the readout path, including the detection loss, leads to a loss of quantum correlations due to mixing with vacuum. We model this loss with a beam splitter of power transmissivity $\eta = 1-\lambda_r$ and reflectivity (loss) $1-\eta = \lambda_r$ which mixes in vacuum $\mathrm{n}$:
\begin{equation}
\mathrm{\tilde{b}}(\Omega) = \sqrt{\eta}\mathrm{b}(\Omega) + \sqrt{1-\eta} \mathrm{n}
\end{equation}

The balanced homodyne detection on the output $\mathrm{\tilde{b}}$ at homodyne angle $\zeta$ provides the values
\begin{equation}
y(\Omega) = \{\cos \zeta, \sin \zeta\}^{\mathrm T} \mathrm{\tilde{b}}(\Omega) = \mathcal{H}^{\rm T}\mathrm{\tilde{b}}(\Omega)
\end{equation}
\begin{equation}
  y(\Omega) =  \sqrt{\eta}\mathcal{H}^{\rm T} \left( -\mathcal{R} \mathrm{\hat{a}}  + \mathcal{T}\mathrm{\hat{v}} + \mathcal{L}_{b1} \mathrm{\hat{n}_1} + \mathcal{L}_{b2} \mathrm{\hat{n}_2} \right) + \sqrt{\eta}\mathcal{H}^{\rm T}\mathcal{Z} \hat{x}_-(\Omega)  + \sqrt{1-\eta} \mathcal{H}^{\rm T}\mathrm{n}(\Omega)
\end{equation}
which we renormalize to the differential mirror displacement
\begin{equation}
\tilde{y} = \frac{\mathcal{H}^{\rm T} \left( -\mathcal{R} \mathrm{\hat{a}}  + \mathcal{T}\mathrm{\hat{v}} + \mathcal{L}_{b1} \mathrm{\hat{n}_1} + \mathcal{L}_{b2} \mathrm{\hat{n}_2} \right)}{\mathcal{H}^{\rm T}\mathcal{Z}}  + \frac{ \sqrt{1-\eta} \mathcal{H}^{\rm T}\mathrm{n} }{ \sqrt{\eta}\mathcal{H}^{\rm T}\mathcal{Z}} +  \hat{x}_-(\Omega)
\end{equation}

We implement the injection of the squeezing from the outside, by defining an action of the squeezing operation on the input field $\mathrm{\hat{a}}$ as:
\begin{equation}
\mathrm{\hat{a}} = \mathcal{S}_{\rm ext}[\phi_{\rm ext}]\mathrm{\hat{a}}^{\rm vac},
\end{equation}
where $\mathrm{\hat{a}}^{\rm vac}$ is the vacuum field before squeezing, and the squeezing matrix with squeeze factor $q_{\rm ext}$  and squeeze angle $\phi_{\rm ext}$ is defined as
\begin{equation}
\mathcal{S}_{\rm ext}=
O(\phi_{\rm ext}) \{\{e^{q_{\rm ext}},0\},\{0,e^{-q_{\rm ext}}\}\}
O(-\phi_{\rm ext}).
\end{equation}
All other fields $\mathrm{\hat{v}},\mathrm{\hat{n}},\mathrm{\hat{n}_1},\mathrm{\hat{n}_1}$ are in the vacuum state.

From this we get the spectral density for this output
\begin{equation}\label{eq:sd_full}
S_x(\Omega) = S_{xx}(\Omega) + 2 {\rm Re}[\chi^*_{\rm eff}(\Omega) S_{xF}(\Omega)] + |\chi_{\rm eff}(\Omega)|^2 S_{FF}(\Omega),
\end{equation}
where
\begin{equation}\label{eq:sxx}
  S_{xx} = \frac{\mathcal{H}^{\rm T}(\mathcal{R}\mathcal{S}_{\rm ext} \mathcal{S}_{\rm ext}^\dag \mathcal{R}^\dag + \mathcal{T}\mathcal{T}^\dag + \mathcal{L}_{b1}\mathcal{L}_{b1}^\dag + \mathcal{L}_{b2}\mathcal{L}_{b2}^\dag)\mathcal{H}}{|\mathcal{H}^{\rm T}\mathcal{Z}|^2} + \frac{1-\eta}{\eta} \frac{1}{|\mathcal{H}^{\rm T}\mathcal{Z}|^2},
\end{equation}
\begin{eqnarray}
  &S_{FF}&  =  \hbar^2 k^2 (1+R_e^2)^2  \mathrm{E}^\dag \mathcal{D}_e O(\delta_{\rm arm} \tau_{\rm arm}) \left(\mathcal{T}_d \mathcal{S}_{\rm ext}\mathcal{S}_{\rm ext}^\dag \mathcal{T}_d^\dag  + \mathcal{L}_{d1}\mathcal{L}_{d1}^\dag + \mathcal{L}_{d2}\mathcal{L}_{d2}^\dag\right)O^\dag(\delta_{\rm arm} \tau_{\rm arm}) \mathcal{D}_e^\dag \mathrm{E} + \nonumber \\ &+& \hbar^2 k^2 T_e^2 \mathrm{E}^\dag\mathcal{L}_v \mathcal{L}_v^\dag \mathrm{E},
\end{eqnarray}
\begin{eqnarray}
  S_{xF} &=& \frac{\hbar k}{\mathcal{H}^{\rm T}\mathcal{Z}} \left((1+R_e^2)\mathcal{H}^{\rm T} (-\mathcal{R} \mathcal{S}_{\rm ext}\mathcal{S}_{\rm ext}^\dag\mathcal{T}_d^\dag + \mathcal{L}_{b1}\mathcal{L}_{d1}^\dag + \mathcal{L}_{b2}\mathcal{L}_{d2}^\dag) O^\dag(\delta_{\rm arm} \tau_{\rm arm}) \mathcal{D}_e^\dag \mathrm{E} e^{-i\Omega \tau_{\rm arm}} + \right. \nonumber \\ &+& \left. T_e \mathcal{H}^{\rm T}\mathcal{T} \mathcal{L}_v^\dag \mathrm{E} \right).
\end{eqnarray}

Finally we normalize the spectral density to the gravitational-wave strain yielding (taking into account the effects of high-frequency corrections~\cite{Rakhmanov2008})
\begin{equation}
S_h(\Omega) = S_x(\Omega) \frac{4}{m^2 L^2 \Omega^4 |\chi_{\rm eff}(\Omega)|^2} \frac{\sin^2 \Omega \tau_{\rm arm}}{\Omega^2 \tau_{\rm arm}^2}\, .
\end{equation}

\subsection{Filter cavities}
Filter cavities on the can be used to create a necessary frequency dependence of quantum correlations, such that the QRPN is suppressed or evaded completely.
There are two scenarios, input filter cavity, where the injected squeezing becomes frequency dependent, and output filter cavity, where the homodyne detection becomes frequency dependent.
We follow \cite{Danilishin2012} and consider a lossless filter cavity, so that the only effect of the cavity is a frequency-dependent rotation of the input squeezed state $\mathrm{\hat{a}} \rightarrow O[\theta_f(\Omega)]\mathrm{\hat{a}}$ or output $\mathrm{b}(\Omega) \rightarrow O[\theta_f(\Omega)]\mathrm{b}(\Omega)$, by the angle
\begin{equation}
\theta_f(\Omega) = \arctan \frac{2\gamma_f \delta_f}{\gamma_f^2 - \delta_f^2 + \Omega^2},
\end{equation}
where $\gamma_f$ is the filter cavity bandwidth, and $\delta_f$ is it's detuning from resonance.
To obtain the spectral corresponding spectral densities it's sufficient to modify the squeeze angle $\phi_{\rm ext} \rightarrow \phi_{\rm ext} + \theta_f(\Omega)$ or homodyne angle $\zeta \rightarrow \zeta - \theta_f(\Omega)$ in the equations for the spectral density Eq.~\ref{eq:sd_full}.
The optimal detuning is on the slope of the cavity resonance $\delta_f = \gamma_f $, and the exact choice of cavity linewidth depends on the parameters of the detector, including the internal squeezing strength and readout loss.

%\bibliography{combined}
% \input{test_bib.tex}
%\bibliographystyle{Science}

\end{document}